\documentclass[preprint2]{aastex6}
\usepackage{natbib}
\usepackage{color}
\usepackage[english]{babel}
\usepackage[normalem]{ulem}
\usepackage{blindtext}
\usepackage{textgreek}
\usepackage{graphicx}
\newcommand{\etal}{et\,al.}
\newcommand{\halpha}{H$\alpha$}
 
\newcommand{\lsim}{\raise0.3ex\hbox{$<$}\kern-0.75em{\lower0.65ex\hbox{$\sim$}}}
\newcommand{\gsim}{\raise0.3ex\hbox{$>$}\kern-0.75em{\lower0.65ex\hbox{$\sim$}}}
\newcommand{\msun}{M$_{\odot}$}

\newcommand{\kms}{km\,s$^{-1}$}
\newcommand{\farcsec}{\mbox{\ensuremath{.\!\!^{\prime\prime}}}}

\begin{document}
\title{The Enigmatic (Almost) Dark Galaxy Coma\,P:  The Atomic
  Interstellar Medium}

\author{Catherine Ball$^1$, John M. Cannon$^1$, Lukas Leisman$^2$, Elizabeth A.K. Adams$^{3,4}$, Martha P. Haynes$^5$, Gyula I.G. J{\'o}zsa$^{6,7,8}$, Kristen B. W. McQuinn$^9$, John J. Salzer$^{10}$, Samantha Brunker$^{10}$, Riccardo Giovanelli$^5$, Gregory Hallenbeck$^{11}$, William Janesh$^{10}$, Steven Janowiecki$^{12}$, Michael G. Jones$^{13}$, Katherine L. Rhode$^{10}$}
\affil{$^1$Department of Physics \& Astronomy, Macalester College, 
1600 Grand Avenue, Saint Paul, MN 55105, USA; cball@macalester.edu \\ 
$^2$Department of Physics and Astronomy, Valparaiso University, 
Valparaiso, IN 46383\\
$^3$ASTRON, the Netherlands Institute for Radio Astronomy, Postbus
  2, 7990 AA, Dwingeloo, The Netherlands\\
$^4$Kapteyn Astronomical Institute, University of Groningen,
  Postbus 800, 9700 AV Groningen, The Netherlands\\
$^5$Center for Astrophysics and Planetary Science, Space
  Sciences Building, 122 Sciences Drive, Cornell University,
  Ithaca NY 14853 USA\\
$^6$SKA South Africa, Radio Astronomy Research Group, 3rd Floor,
  The Park, Park Road, Pinelands, 7405, South Africa\\
$^7$Rhodes University, Department of Physics and Electronics,
  Rhodes Centre for Radio Astronomy Techniques \& Technologies, PO Box
  94, Grahamstown, 6140, South Africa\\
$^8$Argelander-Institut f{\"u}r Astronomie, Auf dem H{\"u}gel 71, 
53121 Bonn, Germany\\
$^9$University of Texas at Austin, McDonald Observatory, 2515 Speedway, 
Stop C1402, Austin, TX 78712, USA\\  
$^{10}$Department of Astronomy, Indiana University, 727 East
  Third Street, Bloomington, IN 47405, USA\\
$^{11}$Washington \& Jefferson College, 
Department of Computing and Information Studies, 
60 South Lincoln Street, Washington PA 15301, USA\\
$^{12}$International Centre for Radio Astronomy Research,
University of Western Australia,
35 Stirling Highway,
Crawley, WA 6009, Australia\\ 
$^{13}$Instituto de Astrof{\'i}sica de Andaluc{\'i}a,
Glorieta de la Astronomía s/n, 18008 Granada,
Spain}
\begin{abstract}

We present new high-resolution HI spectral line imaging of Coma\,P,
the brightest HI source in the system HI\,1232$+$20.  This galaxy with extremely
low surface brightness was first identified in the ALFALFA
survey as an ``(Almost) Dark'' object: a clearly extragalactic HI
source with no obvious optical counterpart in existing optical survey
data (although faint ultraviolet emission was detected in archival
GALEX imaging).  Using a combination of data from the Westerbork
Synthesis Radio Telescope and the Karl G. Jansky Very Large
Array, we investigate the HI morphology and kinematics at a
variety of physical scales.  The HI morphology is irregular, reaching
only moderate maxima in mass surface density (peak $\sigma_{\rm HI}$
$\sim$ 10 \msun\,pc$^{-2}$). Gas of lower surface brightness extends to
large radial distances, with the HI diameter measured at 4.0$\pm$0.2
kpc inside the 1 \msun\,pc$^{-2}$ level.  We quantify the
relationships between HI gas mass surface density and star formation
on timescales of $\sim$100-200 Myr as traced by GALEX far ultraviolet
emission. While Coma\,P has regions of dense HI gas reaching the
N$_{\rm HI}$ $=$ 10$^{21}$ cm$^{-2}$ level typically associated with
ongoing star formation, it lacks massive star formation as traced by
H$\alpha$ emission.  The HI kinematics are extremely complex: a simple
model of a rotating disk cannot describe the HI gas in Coma\,P.  Using
spatially resolved position-velocity analysis we identify two nearly
perpendicular axes of projected rotation that we interpret as either
the collision of two HI disks or a significant infall event.
Similarly, three-dimensional modeling of the HI dynamics provides a
best fit with two HI components. Coma\,P is just consistent (within
3\,$\sigma$) with the known M$_{\rm HI}$ -- D$_{\rm HI}$ scaling
relation.  It is either too large for its HI mass, has too low an HI
mass for its HI size, or the two HI components artificially extend its
HI size.  Coma\,P lies within the empirical scatter at the faint end
of the baryonic Tully--Fisher relation, although the complexity of the
HI dynamics complicates the interpretation.  Along with its large ratio of HI
to stellar mass, the collective HI characteristics of Coma P
make it unusual among known galaxies in the nearby universe.

\end{abstract}						

\keywords{galaxies: evolution --- galaxies: dwarf --- galaxies:
  irregular --- galaxies: individual (Coma\,P, AGC\,229385)}

\section{Introduction}
\label{S1}
  
The ALFALFA blind HI extragalactic survey has now produced the largest
and most diverse catalog of HI-detected systems ever assembled
\citep{giovanelli05,haynes11}. The final source catalog contains more
than 31,500 extragalactic sources of high significance covering over 7
dex in HI mass.  The ALFALFA catalog provides unique perspectives on
the population of gas-rich extragalactic sources.

Our group has been actively investigating the physical properties of
optically faint galaxies as the ALFALFA catalog has matured and neared
completion. These efforts have included both interferometric HI
spectral line imaging as well as deep ground-based imaging of selected
sources.  We have targeted both systems that are candidate Local
Group/Volume objects and systems that are almost certainly well
outside the Local Group.

This first line of inquiry has led to the characterization of extreme
regions of extragalactic parameter space. In the ``Survey of HI in
Extremely Low-mass Dwarfs'' \citep[hereafter ``SHIELD'';][]{cannon11a}
we are studying those ALFALFA-detected galaxies with HI mass
reservoirs smaller than 10$^{7.2}$ \msun.  Using Hubble Space
Telescope (HST) imaging, \citet{mcquinn15a} show that these galaxies
are characterized by fluctuating, non-deterministic, and inefficient
star formation.  The nature of the star formation process in these
galaxies is studied in further detail in \citet{teich16}, where it is
shown that stochasticity plays an important role in governing the
relationships between recent star formation and neutral hydrogen gas.
A companion paper \citep{mcnichols16} explores the rotational dynamics
of these sources, highlighting that galaxies in this mass range probe
the transition regime from rotational to pressure support.

This line of inquiry has also led to the ALFALFA discovery of some of
the most extreme Local Volume (D $\lsim$ 11 Mpc) galaxies known. A
series of manuscripts describes Leo\,P, an extremely low-mass galaxy
just outside the Local Group with a single HII region and an extremely
low oxygen abundance \citep{giovanelli13, mcquinn13, mcquinn15b,mcquinn15c,rhode13, skillman13,ezbc14,warren15}.  Subsequently,
the even more metal-deficient Leoncino Dwarf (AGC\,198691) was
discovered by ALFALFA \citep{hirschauer16}.

Moreover, follow-up optical and HI observations have also produced
multiple detections of resolved stellar populations in ``ultra-compact
high velocity clouds'' (UCHVCs). These are extended HI systems that
have properties such as compact, low-mass halos if they are indeed nearby
(within $\sim$1-2 Mpc). We refer the reader to
\citet{adams13,adams15,adams16} and \citet{janesh15,janesh17} for
details about the population of these nearby systems.

The second line of inquiry targets optically faint objects that appear
to be outside of the Local Volume.  This includes a very small subset
of ALFALFA HI detections ($<$1\%) that have clear optical counterparts,
that have velocities that indicate they are clearly extragalactic, and
that are not obviously tidal debris. We have labeled these objects as
``(Almost) Dark'' galaxies. These enigmatic sources push the
boundaries of what might typically be labeled as a ``galaxy''; they
thus offer unique and powerful insights into the extrema of various
fundamental scaling relations.

In \citet{cannon15} we presented the basic selection criteria that are
used to identify clearly extragalactic (i.e., with recessional
velocities sufficient to cleanly separate them from Milky Way HI gas)
sources that are (Almost) Dark galaxy candidates.  In short, such
systems lack an obvious optical counterpart in Sloan Digital Sky Survey (SDSS) imaging (which can
be a result of intrinsically low surface brightness or of spatial
coincidence) and are unlikely to be tidal in origin.  These systems
naturally have elevated M$_{\rm HI}$/L$_{\rm B}$ ratios
(\citet{cannon15} note that empirically the sources usually have
M$_{\rm HI}$/L$_{\rm B}$ $\gsim$ 10).  As discussed in
\citet{leismanthesis} and \citet{leisman17}, roughly 1\% of the
sources in the ALFALFA catalog meet these criteria and are worthy of
detailed further exploration to determine their physical properties.
\begin{figure*}
	\begin{center}
		\includegraphics[width=0.75\textwidth]{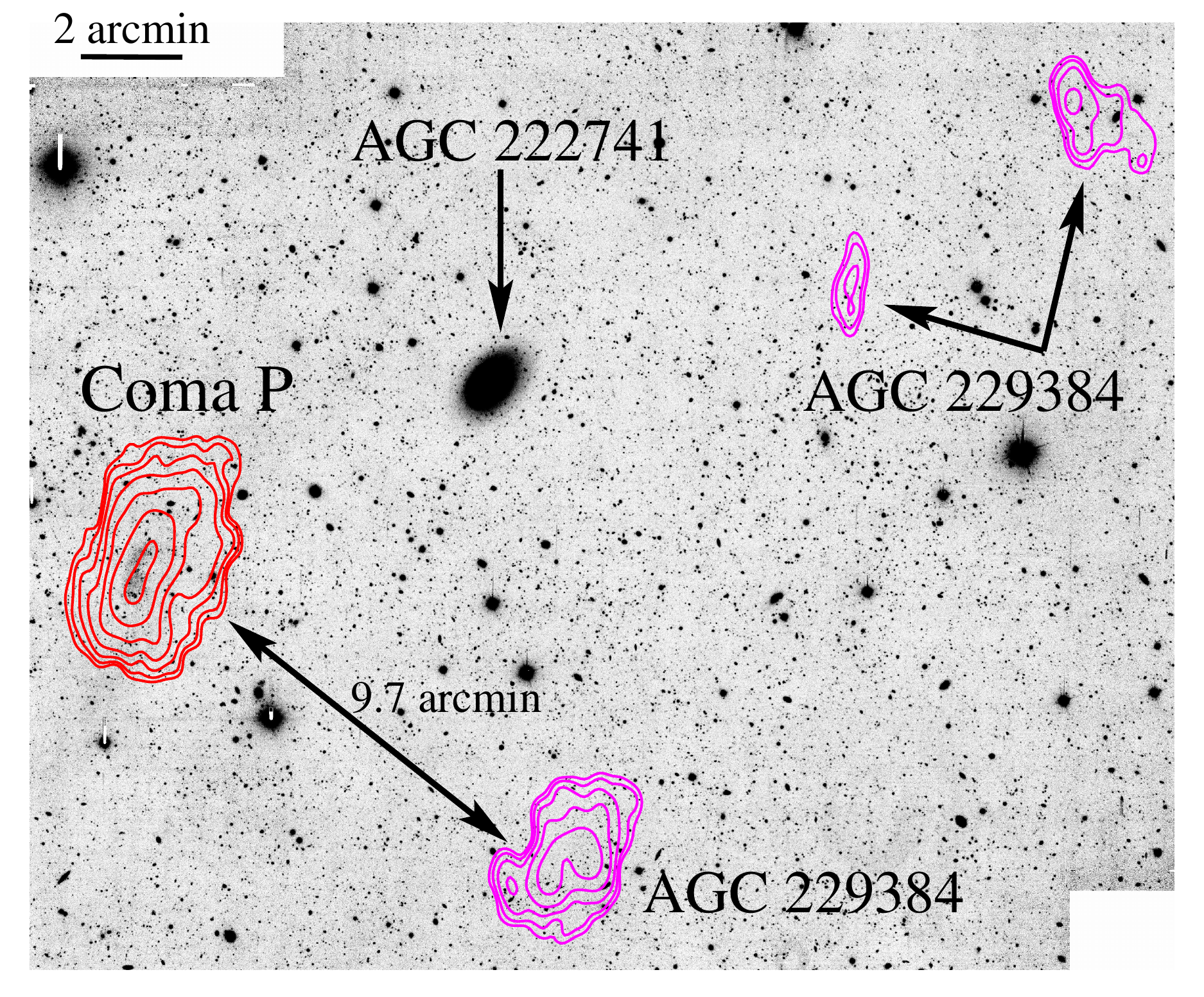}
	\end{center}
	\clearpage
	\caption{pODI image of the Coma P field, overlaid with contours of HI column density from WSRT imaging as presented in \citet{janowiecki15}.  Coma P is located in close angular and velocity proximity to two other HI line sources identified in the ALFALFA survey.  Coma P (V$_{\rm HI}$ $=$ 1342.6 $\pm$ 1 \kms) is separated by $\sim$9.\arcmin 7 from AGC 229384 (V$_{\rm HI}$ $=$ 1309 $\pm$ 1 \kms) and by $\sim$14.\arcmin 4 from the eastern component of AGC 229383 (V$_{\rm HI}$ $=$ 1282 $\pm$ 1 \kms). Although the recessional velocity of Coma P is large, the HST images presented in S. Brunker \etal\ (2017, in preparation) reveal it to be located at D $=$ 5.50 $\pm$ 0.28 Mpc.  The distances of AGC 229384 and AGC 229383 remain uncertain, and whether they are physically associated with Coma P is not known with present data. AGC 222741 is also an HI line source but with a significantly higher HI recessional velocity (V$_{\rm HI}$ $=$ 1884 $\pm$ 2 \kms).  The WSRT contours show HI column densities spaced logarithmically between 1 and 64 $\times$ 10$^{19}$ cm$^{-2}$. The WSRT beam (13\arcsec $\times$ 39\arcsec) is elongated north--south.}
	\label{ComaPfield}
\end{figure*}
Characterizing these (Almost) Dark galaxies requires intensive follow-up observations, including HI synthesis observations and deep
ground-based optical imaging. Previous work by our group has shown
that, to date, all of the candidate (Almost) Dark objects turn out to
not be truly dark galaxies.  Follow-up observations reveal complex
tidal systems \citep{leisman16}, peculiar sources with large pointing
offsets between the ALFALFA and the interferometric centroids
\citep{cannon15,singer17}, galaxies with low surface brightness similar to
ultradiffuse galaxies \citep{leisman17}, and sources with other
interpretive issues \citep[see][]{leismanthesis}.

The source that is the focus of this article lies at the
intersection of these two efforts. Drawn from the population of more
distant (Almost) Dark sources, recent revisions to its distance show
that it is actually quite nearby - an extreme Local Volume source that still
pushes typical definitions of ``galaxy.'' This particular source
attracted attention as being completely undetected in SDSS, but detected
with a signal-to-noise ratio S/N $>$ 99 in ALFALFA data products. Examining archival GALEX
imaging of the field suggested an optical counterpart. Subsequent
detailed HI and optical follow-up observations presented in
\citet{janowiecki15} revealed a stellar population of extremely low surface brightness
stellar population with extended HI gas. This source, AGC 229385, we
hereafter will refer to as ``Coma P.''

Figure~\ref{ComaPfield} shows the same comparison of the optical and
HI morphologies of Coma\,P as shown in \citet{janowiecki15}.  This
galaxy is part of a trio of HI line sources with similar recessional
velocities that were identified by ALFALFA and confirmed in the
Westerbork Synthesis Radio Telescope (WSRT\footnote{The Westerbork
  Synthesis Radio Telescope is operated by ASTRON, the Netherlands
  Institute for Radio Astronomy, with support from the Netherlands
  Foundation for Scientific Research (NWO).})  HI images presented in
\citet{janowiecki15}.  Very deep optical images acquired with the
partially populated One Degree Imager (pODI) on the WIYN\footnote{The
  WIYN Observatory is a joint facility of the University of
  Wisconsin-Madison, Indiana University, the University of Missouri,
  and the National Optical Astronomy Observatory.}  3.5\,m telescope
at Kitt Peak Observatory\footnote{Kitt Peak National Observatory,
  National Optical Astronomy Observatory, which is operated by the
  Association of Universities for Research in Astronomy (AURA) under
  cooperative agreement with the National Science Foundation.}
revealed an extended stellar component of low surface brightness
associated with the brightest HI source (Coma\,P).  Stellar
counterparts of the other two HI sources (AGC\,229384 and AGC\,229383)
remain undetected, with upper limits on the surface brightness of 27.9 and
27.8 mag arcsec$^{-2}$ in the g$^{\prime}$ filter, respectively.
These sources have similar systemic velocities to Coma\,P (V$_{\rm
  HI}$ $=$ 1309\,$\pm$\,1 \kms\ and 1282\,$\pm$\,4 \kms\ for
AGC\,229384 and AGC\,229383, respectively) and are offset by
$\sim$9\arcmin\.7 and $\sim$15\arcmin\ from Coma\,P (see
Figure~\ref{ComaPfield}).  \citet{janowiecki15} discuss the physical
properties of these sources, based on their adopted distance of 25
Mpc.

Although the recessional velocity of Coma\,P \citep[V$_{\rm HI}$ $=$
  1348\,$\pm$\,1 \kms;][]{janowiecki15} and Virgocentric flow model
\citep{masters05} for this region of the sky suggest that it is
located well beyond the Virgo Cluster, recent HST images of the source
have demonstrated that it is located significantly closer to us.  As
discussed in detail in S. Brunker \etal\ (2017, in preparation), the red giant
branch of the HST color magnitude diagram is best fit by a distance of
only 5.50$\pm$0.28 Mpc.  The physical properties of Coma\,P, which are
summarized in Table~\ref{t1}, become increasingly extreme as a result
of this change in distance.  We note that the HST images do not cover
either of the other HI sources that are detected in the ALFALFA and
WSRT data as shown in \citet{janowiecki15} and
Figure~\ref{ComaPfield}.  While their proximity in both space and
velocity strongly suggests association, we do not know whether these objects
are at the same distance as Coma\,P or not.

In this work, we present new HI line synthesis observations of the
neutral interstellar medium of Coma\,P constructed by combining new
higher resolution Very Large Array (VLA\footnote{The National Radio
  Astronomy Observatory is a facility of the National Science
  Foundation operated under cooperative agreement by Associated
  Universities, Inc.}) HI line imaging with the WSRT HI line dataset
presented in \citet{janowiecki15}. The resultant spectral grids are
used to study the morphology and dynamics of the HI gas at a variety
of angular resolutions.  The higher resolution of this new combined
dataset allows detailed kinematic and morphological analysis of the HI
gas in Coma\,P, which we present below.  We also leverage the
spatially resolved nature of the stellar population from the HST
images in order to investigate the star formation process in this
enigmatic galaxy.

This paper is organized as follows.  In \S~\ref{S2} we discuss the HI
data handling and combination.  \S~\ref{S3} contains detailed
discussion about the HI morphology and kinematics.  In \S~\ref{S4} we
compare the HI properties with those of the stellar populations.  The
HI dynamics of Coma\,P are studied and modeled in \S~\ref{S5}.  We
contextualize Coma\,P in \S~\ref{S6} and summarize our conclusions in
\S~\ref{S7}.

\begin{deluxetable}{lcc}  
	\tablecaption{Physical Characteristics of Coma\,P\label{t1}} 
	\tablewidth{0pt}  
	\tablehead{ 
		\colhead{Parameter} &\colhead{Value}}    
	\startdata       
	R.A. (J2000)                         &12$^{\rm h}$ 32$^{\rm m}$ 10$^{\rm s}$.3\\        
	Decl. (J2000)                             &$+$20\arcdeg 25\arcmin 23\arcsec\\    
	Adopted distance (Mpc)                          &5.50\,$\pm$\,0.28\tablenotemark{a}\\
	M$_{\rm B}$ (mag)                               &$-$10.71$\pm$0.14\tablenotemark{a}\\
	M$_{\star}$ (\msun)                              &(1.0$\pm$0.3)\,$\times$\,10$^{6}$\tablenotemark{a}\\
	M$_{\rm HI}$ (\msun)                             &(3.48$\pm$\,0.35)\,$\times$\,10$^{7}$\tablenotemark{b}\\
	D$_{\rm HI}$ (kpc)\tablenotemark{c}              &4.0$\pm$0.2\\ 
	SFR$_{\rm FUV}$ (\msun\,yr$^{-1}$)                &(3.1$\pm$1.8)\,$\times$\,10$^{-4}$ 
	\enddata     
	\tablenotetext{a}{S. Brunker \etal\ (2017, in preparation)} 
	\tablenotetext{b}{Using S$_{\rm HI}$ $=$ 4.87\,$\pm$\,0.04 Jy\,km\,s$^{-1}$ from Haynes \etal\ (2011).}
	\tablenotetext{c}{Measured at the 1 \msun\,pc$^{-2}$ level.}
\end{deluxetable}

\section{Observations and Data Handling}
\label{S2}
\subsection{HI Data Products}
\label{S2.1}

\begin{figure*}
	\vspace{-0.5 cm}
	\begin{center}
		\includegraphics[width=0.85\textwidth]{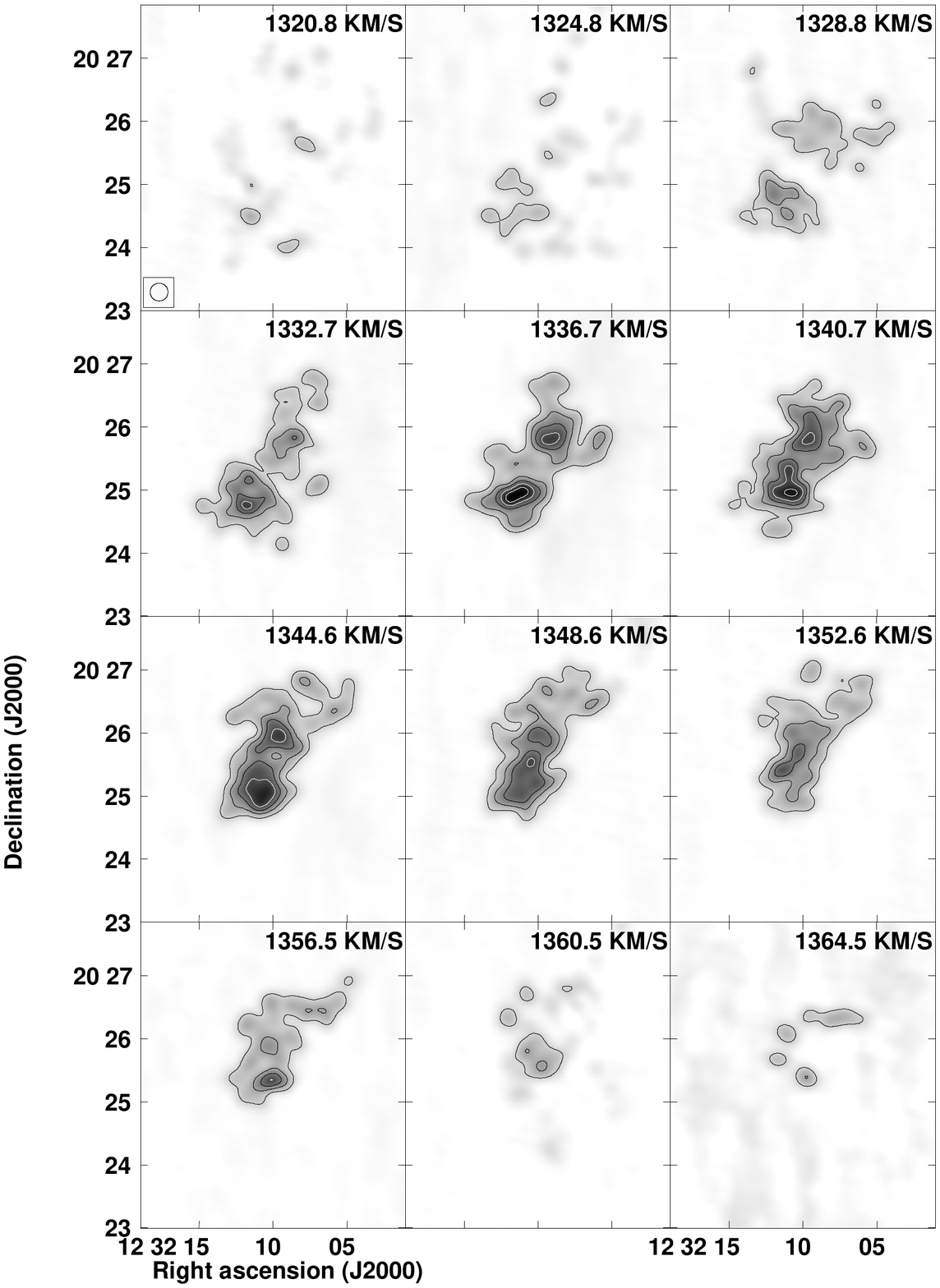}
	\end{center}
	\vspace{-1.5 cm}
	\caption{Channel maps showing the low-resolution (17\arcsec\ beam)
		three-dimensional VLA$+$WSRT HI data cube of Coma\,P.  Contours are
		overlaid at (3,6,9,12,15)\,$\sigma$, where $\sigma$ $=$ 0.66 mJy\,beam$^{-1}$ in
		line-free channels of the final data cube (see Table~\ref{t2}).}
	\label{CHMAPS.LOW}
\end{figure*}

\begin{figure*}
	\vspace{-0.5 cm}
	\begin{center}
		\includegraphics[width=0.9\textwidth]{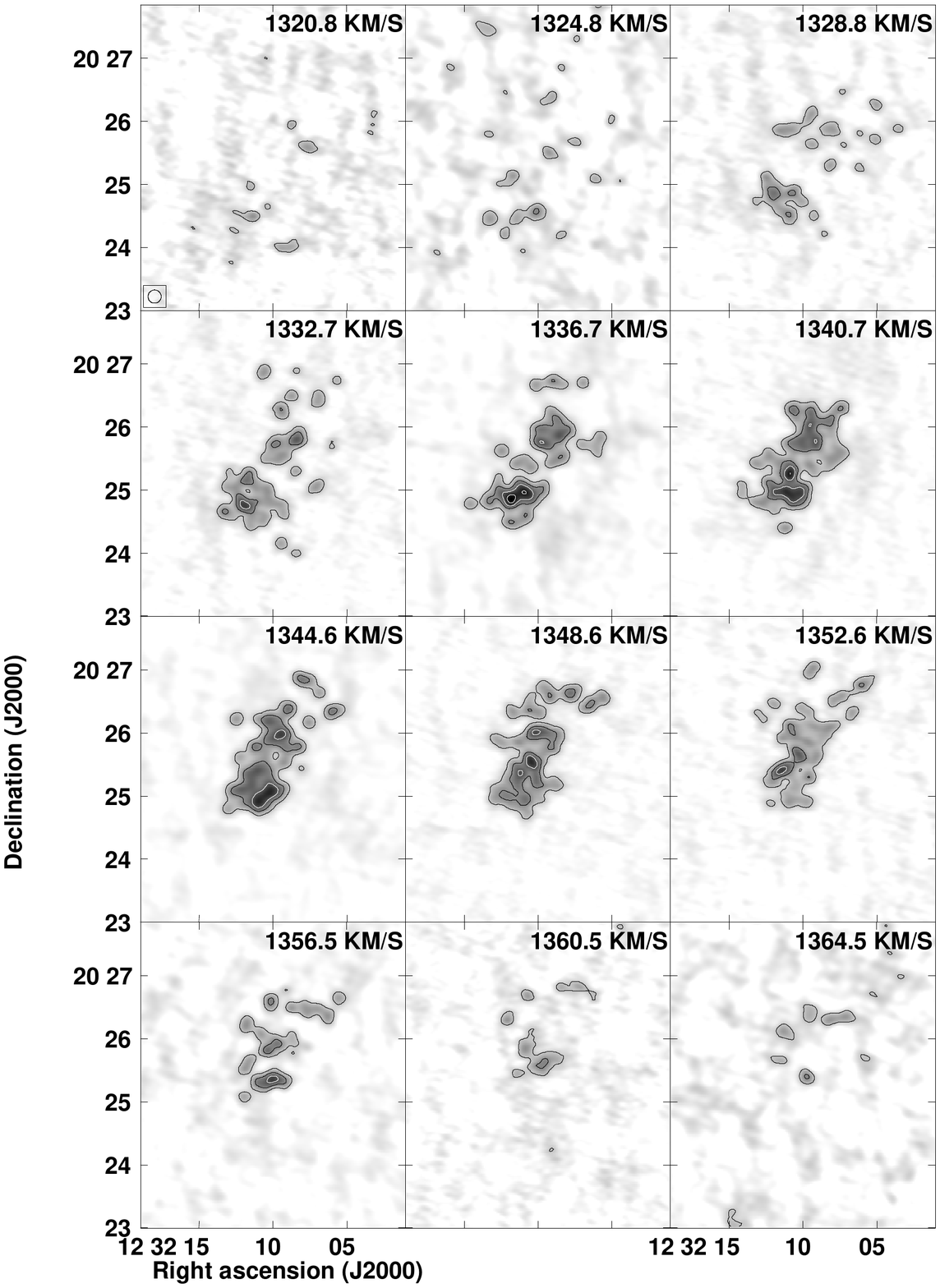}
	\end{center}
	\vspace{-1.5 cm}
	\caption{Channel maps showing the medium-resolution (12\farcsec5 beam)
		three-dimensional VLA$+$WSRT HI data cube of Coma\,P.  Contours are
		overlaid at (3,6,9,12)\,$\sigma$, where $\sigma$ $=$ 0.56 mJy\,beam$^{-1}$ in
		line-free channels of the final data cube (see Table~\ref{t2}).}
	\label{CHMAPS.MED}
\end{figure*}

\begin{figure*}
	\vspace{-0.5 cm}
	\begin{center}
		\includegraphics[width=0.9\textwidth]{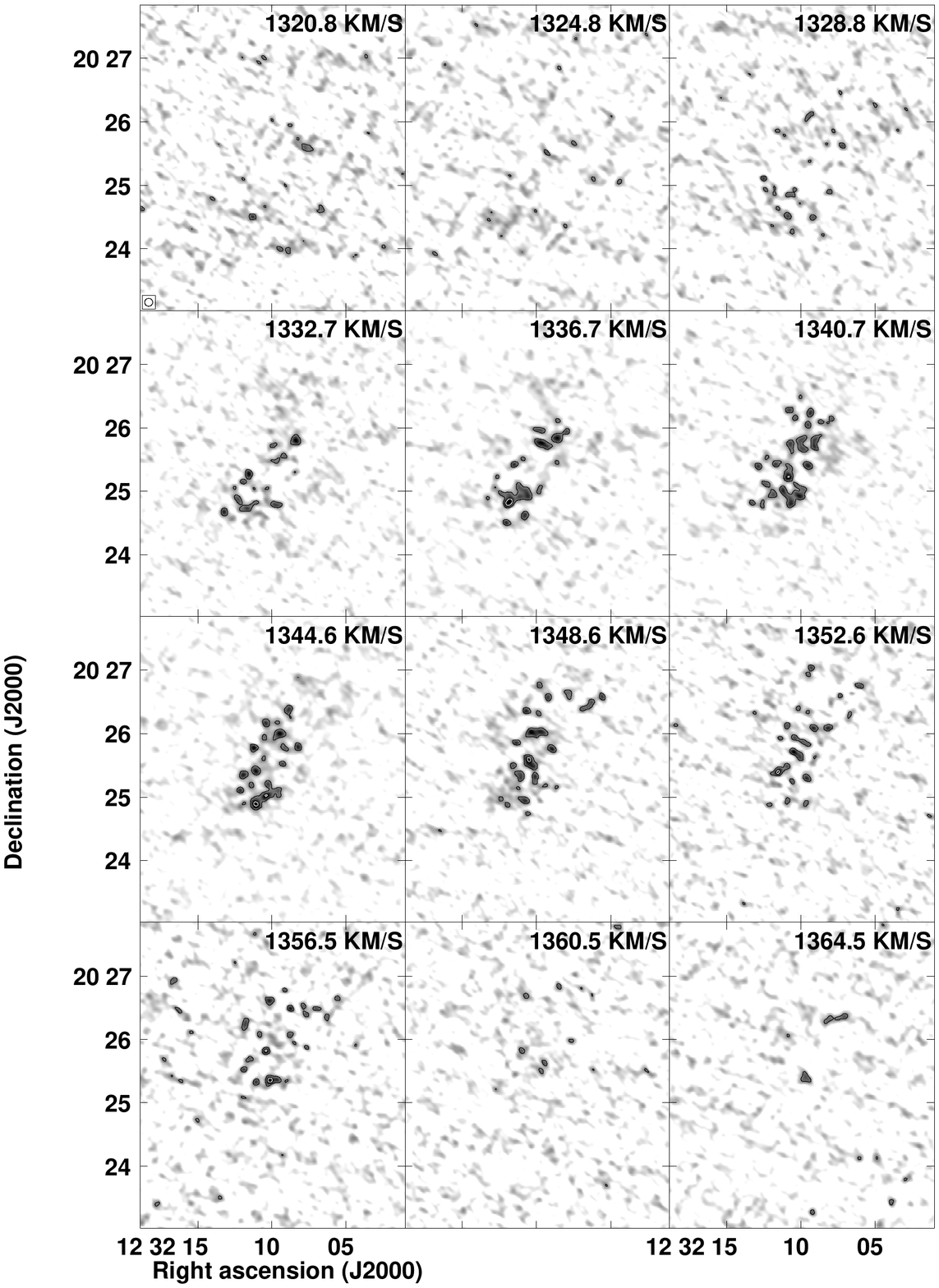}
	\end{center}
	\vspace{-1.5 cm}
	\caption{Channel maps showing the high-resolution (7\farcsec 5 beam)
		three-dimensional VLA$+$WSRT HI data cube of Coma\,P.  Contours are
		overlaid at (3,6)\,$\sigma$, where $\sigma$ $=$ 0.69 mJy\,Bm$^{-1}$ in
		line-free channels of the final data cube (see Table~\ref{t2}).}
	\label{CHMAPS.HIGH}
\end{figure*}
We use the same WSRT data products as those presented in
\citet{janowiecki15}, to which we refer the reader for details of the
data handling.  Briefly, three separate 12 hr synthesis tracks were
obtained for program R13B/001 (P.I. Adams).  A 10 MHz wide band
is separated into 1024 channels, delivering a native spectral
resolution of 2.06 \kms\,ch$^{-1}$.  Following manual excision of
radio-frequency interference and bad data, the bandpass shape was
removed using observations of the primary calibrator.  The data were
then self-calibrated using a continuum image of the target field.

VLA HI spectral line imaging of Coma\,P was obtained in 2015 February and
March for program VLA/15A-307 (P.I. Leisman).  A total of 8
hr of observing time was acquired in the ``B'' configuration of the
array (maximum and minimum baseline lengths of $\sim$53 and 1
k$\lambda$ at 21 cm, nominal beam size $\sim$6\arcsec).  The primary
calibrator was 3C286 and the phase calibrator was J1158$+$2450.
Observational overheads resulted in a total of 6.3 hr of on-source
integration time.  Reduction of the VLA data followed standard
practices within the environment of the Common Astronomy Software Application 
\citep[CASA\footnote{https://casa.nrao.edu/};][]{mcmullin07}.  

\subsection{Combination Imaging}
\label{S2.2}

In order to leverage both the surface brightness sensitivity of the
WSRT data and the angular resolution of the VLA data, we use both sets
of calibrated visibilities to create three-dimensional data cubes of HI
line emission.  While the combination of single-dish and
interferometric data is common (the so-called ``zero-spacing
correction''), the combination of datasets from different
interferometric observatories is less frequently employed.  In
addition to the different synthesized beam shapes of the individual
observatories, they also typically employ different weighting schemes
for calibrated visibilities.  We refer the reader to
\citet{carignan98} for an early discussion of techniques.

There are effectively two strategies for the combination of
interferometric datasets.  In the first, appropriately weighted $uv$
visibilities are combined in the image plane
\citep[e.g.,][]{sorgho17}. In the second, the visibilities are
weighted, combined in the $uv$ plane, and imaged by deconvolving with
a beam that combines the shape of the beams for the input datasets
\citep[e.g.,][]{ezbc14}. We favor the second strategy for image
combination, since the CASA CLEAN algorithm is specifically designed
to be able to image multiple datasets on the fly.

The relative weights of the individual data sets were calculated by
measuring the rms as a function of time and of baseline in line-free
channels using the CASA task \textsc{statwt}.  We then performed
rigorous checking for matching HI fluxes per channel between datasets
by imaging the individual, calibrated databases separately and by
enforcing appropriate tapering in the $uv$ domain to achieve similar
beam sizes.  As expected, the full-resolution VLA images recover less
flux than the WSRT images, and both recover less than the ALFALFA flux
(see below).  After tapering to common resolutions we find agreement
between the fluxes per channel in all datasets.

\begin{deluxetable*}{lcccc}  
	\tablecaption{Details of Final HI Mapping Data Products\label{t2}} 
	\tablewidth{0pt}  
	\tablehead{ 
		\colhead{Data Product} &\colhead{Beam Size}            &\colhead{rms}     &\colhead{Peak N$_{\rm HI}$\tablenotemark{a}}       &\colhead{S$_{\rm HI}$} \\ \colhead{}	&\colhead{}	&\colhead{(mJy Bm$^{-1}$)}	&\colhead{($10^{20}$ cm$^{-2}$)}	&\colhead{(Jy km s$^{-1}$)}}
	\startdata       
	VLA data only                    &6\farcsec 90$\times$\,4\farcsec 96 &0.77 &17.25 &3.02\,$\pm$\,0.30\\
	Combined data, High resolution   &7\farcsec 5$\times$\,7\farcsec 5   &0.69 &12.8 &2.92\,$\pm$\,0.29\\
	Combined data, Medium resolution &12\farcsec 5$\times$\,12\farcsec 5 &0.56 &9.3 &2.96\,$\pm$\,0.30\\
	Combined data, Low resolution    &17\arcsec$\times$\,17\arcsec     &0.66 &8.5	&3.56\,$\pm$\,0.36\\
	WSRT data only                   &62\farcsec 5$\times$\,17\farcsec 8 &0.58 &5.89 &4.43\,$\pm$\,0.44 \tablenotemark{b}
	\enddata   
	\tablenotetext{a}{HI column density as defined assumes a uniform source that fills the beam, so it should always increase with \\increasing angular resolution if there is structure.}
	\tablenotetext{b}{Janowiecki \etal\ (2015)}  
	
\end{deluxetable*}
\begin{figure*}
	\includegraphics[width=\textwidth]{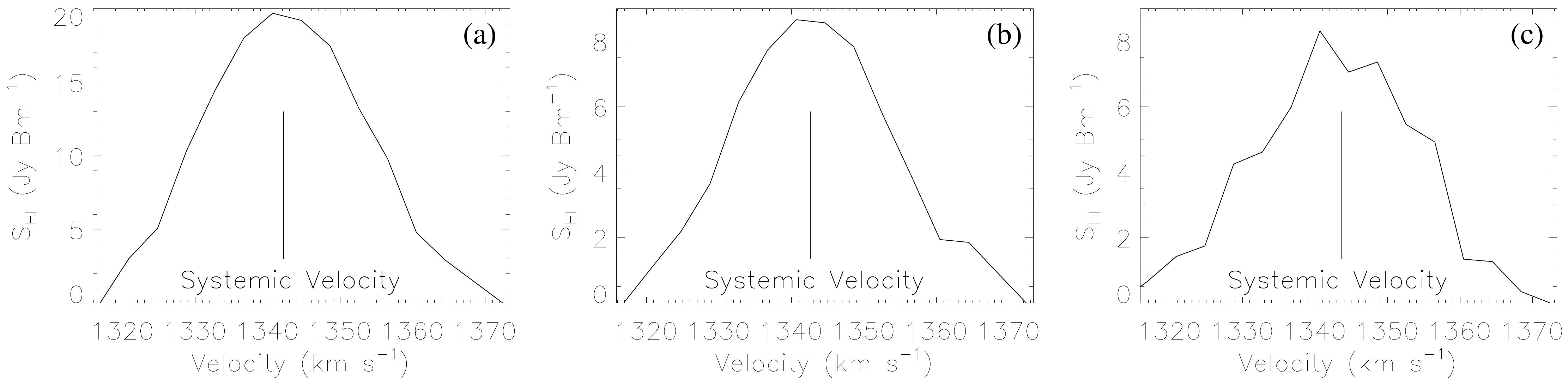}
	\vspace{-0.5 cm}
	\caption{Global HI profiles of Coma\,P at (a) low (17\arcsec\ beam), (b)
		medium (12\farcsec5 beam), and (c) high (7\farcsec 5 beam)
		angular resolutions.  The intensity-weighted mean HI velocity in
		each cube is denoted by a vertical bar. All systemic velocities
		agree within 1 \kms.  We adopt V$_{\rm HI}$ $=$ 1342.6\,$\pm$\,1
		\kms, derived from the medium-resolution data cube, as the systemic
		velocity of Coma\,P.}
	\label{globalprofiles}
\end{figure*}
\begin{figure*}
	\begin{center}
		\includegraphics[width=0.725\textwidth]{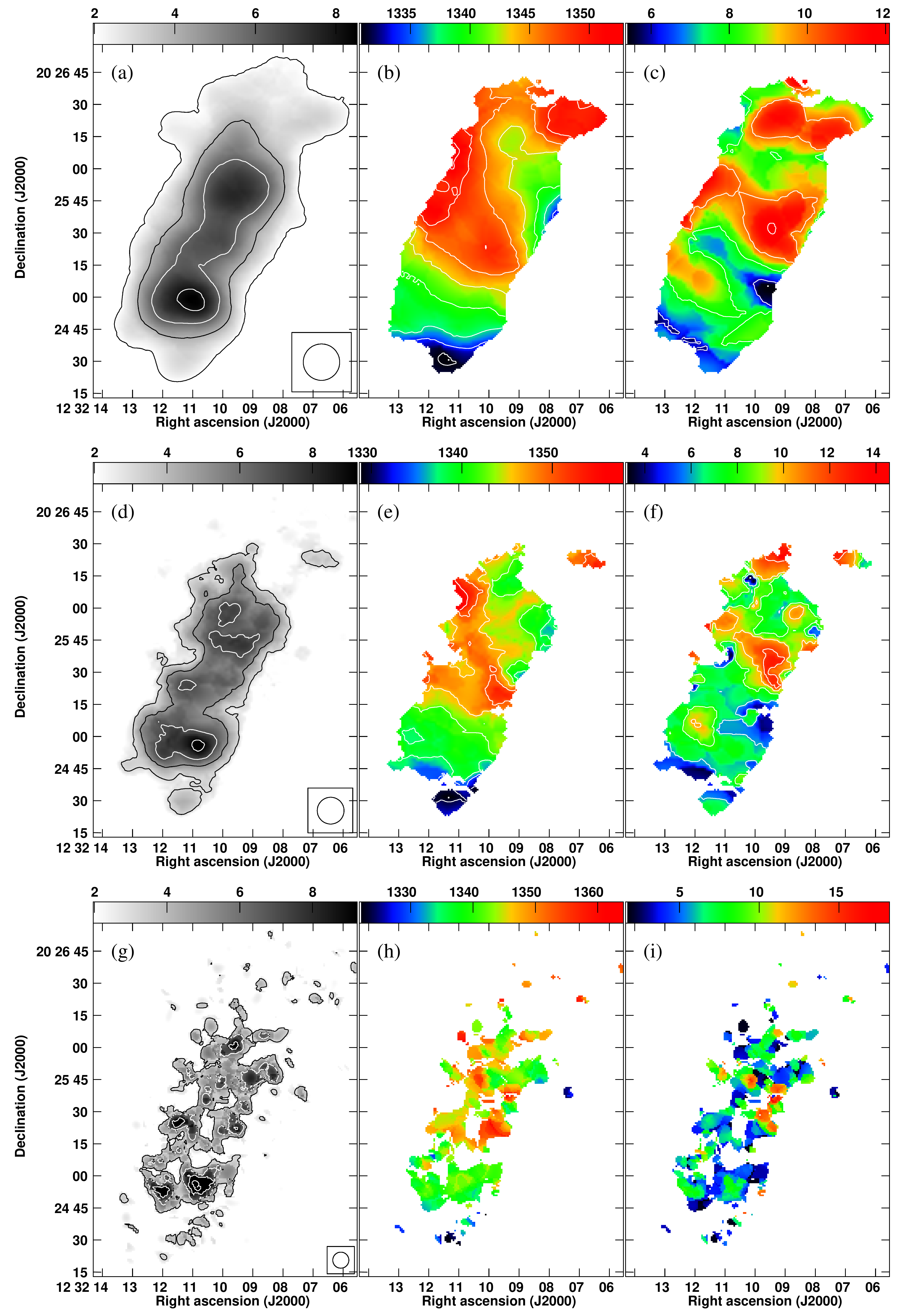}
	\end{center}
	\vspace{-0.5 cm}
	\caption{HI images of Coma\,P at low (top row,
		17\arcsec\ beam), medium (middle row, 12\farcsec 5 beam), and high
		(bottom row, 7\farcsec 5 beam) resolution.  Panels (a), (d), and
		(g) show HI column density in units of 10$^{20}$ cm$^{-2}$.
		Contours are overlaid at levels of (2,4,6,8)\,$\times$\,10$^{20}$
		cm$^{-2}$, (3,5,7,9)\,$\times$\,10$^{20}$ cm$^{-2}$, and
		(3,6,9,12)\,$\times$\,10$^{20}$ cm$^{-2}$ in panels (a), (d), and
		(g), respectively.  Panels (b), (e), and (h) show the
		intensity-weighted HI velocity field in units of \kms. The
		contours in panel (b) span the range of 1331--1352 \kms\ in
		intervals of 3 \kms, while the contours in panel (e) span the range of
		1330--1354 \kms in intervals of 4 \kms.  Panels (c), (f), and (i)
		show the intensity-weighted HI velocity dispersion in units of
		\kms: the contours in panels (c) and (f) are at levels of
		(6,8,10,12) \kms.  No contours are shown in panels (h) and (i) for ease of
		interpretation.}
	\label{HIimages}
\end{figure*}
With the individual $uv$ databases appropriately weighted, we then
imaged all of them simultaneously in the CASA CLEAN algorithm.  We
employ the ``Briggs'' weighting scheme and vary the ``robust''
weighting parameter (which ranges from $-$2 to $+$2 for uniform to
natural weighting) and the $uv$-domain tapering to produce final cubes
at high, medium, and low angular resolutions.  Table~\ref{t2}
summarizes basic parameters of each final dataset.  The final data
cubes were produced using interactive clean boxes and were cleaned to
the 0.5$\sigma$ level (determined in line-free channels of cubes with
very small numbers of clean iterations).  Residual flux rescaling was
employed on each resulting cube following \citet{jorsater95}. We
present the HI data cubes at low, medium, and high resolution in channel
map format in Figures~\ref{CHMAPS.LOW}, \ref{CHMAPS.MED},
\ref{CHMAPS.HIGH}, respectively.

Each resulting data cube was threshold-blanked at the 2$\sigma$ level
and then closely examined by hand for spectral and spatial continuity
of features.  The global HI line emission in these blanked cubes is
plotted in Figure~\ref{globalprofiles}.  The systemic velocity of
Coma\,P in each cube is assigned to the intensity-weighted mean.  The
resulting blanked data cubes were then collapsed to create
two-dimensional images of HI mass surface density, HI intensity-weighted
velocity, and HI intensity-weighted velocity dispersion.  Note by
examining Table~\ref{t2} that the integrated HI flux from Coma\,P
increases as the angular resolution decreases, and that the products with the lowest
angular resolution (WSRT data only, highest surface
brightness sensitivity) recover S$_{\rm HI}$ = 4.43\,$\pm$\,0.44
Jy\,km\,s$^{-1}$.  This agrees with the ALFALFA HI flux integral
(S$_{\rm HI}$ 4.87\,$\pm$\,0.04 Jy\,km\,s$^{-1}$) within errors.

\section{The HI Content of Coma\,P}
\label{S3}
\begin{figure*}[!ht]
	\begin{center}
		\includegraphics[width=0.85\textwidth]{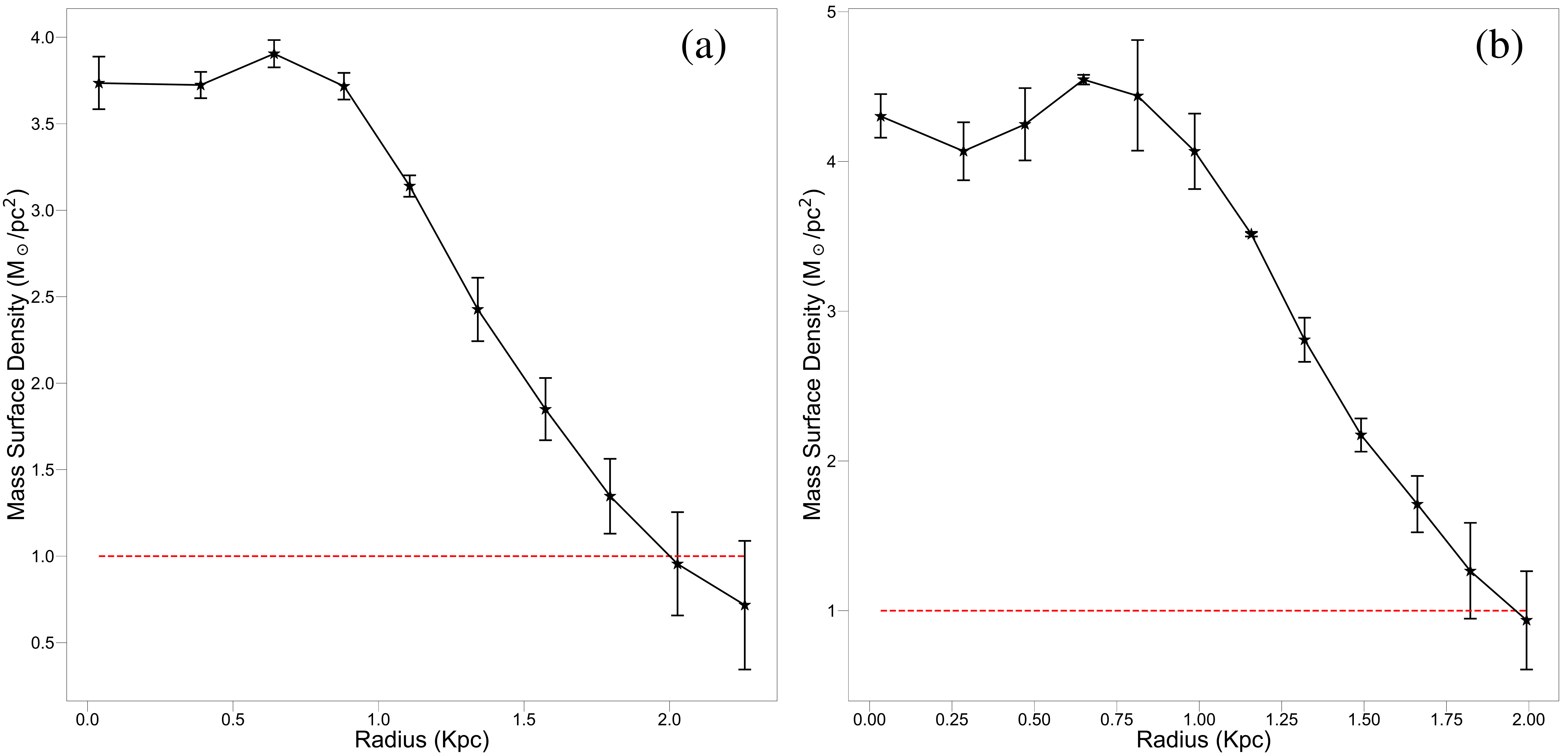}
	\end{center}
	\caption{Radially integrated profiles of HI mass surface density as a
		function of radial distance from the adopted morphological center of
		the galaxy ($\alpha$,$\delta$) = (12$^{\rm h}$32$^{\rm m}$10.$^{\rm
			s}$3, $+$20\arcdeg25$^{\prime}$23$^{\prime\prime}$) using (a) the low-resolution 
		(17\arcsec\ beam or 453 pc physical resolution)
		and (b) the medium resolution (12\farcsec 5 beam or 333 pc physical
		resolution) HI data.  These profiles have been corrected
		for inclination ($i$ $=$ 63\arcdeg) and for beam smearing as
		described in \citet{wang16}.  The dotted red line denotes a mass
		surface density of 1 \msun\,pc$^{-2}$. The HI diameter of Coma\,P is
		4.0\,$\pm$\,0.2 kpc.}\label{HIprofile}
\end{figure*}

Our combined VLA$+$WSRT images reveal the properties of the HI gas in
Coma\,P in unprecedented detail.  Our low-resolution images are in
excellent agreement with those presented in \citet{janowiecki15}.  We
refer the reader to that work for a discussion of the gas with extended low
surface brightness in Coma\,P, as well for details about the other
HI line sources that are in close angular and velocity proximity.

\subsection{HI Morphology}
\label{S3.1}

In Figure~\ref{HIimages} we show two-dimensional representations of
the HI column density (calibrated in units of 10$^{20}$ cm$^{-2}$),
the intensity-weighted HI velocity, and the intensity weighted HI
velocity dispersion.  Each of these images is shown at low
(17\arcsec\ beam), medium (12\farcsec 5 beam), and high
(7\farcsec 5 beam) angular resolutions.  The corresponding elements of physical
resolution are 453 pc, 333 pc, and 200 pc, respectively,
assuming the distance of 5.5 Mpc from S. Brunker \etal\ (2017, in preparation).

In panels (a), (d), and (g) of Figure~\ref{HIimages}, the HI column
density distributions are shown in grayscale and with uniformly spaced
overlying contours.  At $\sim$450 pc resolution the HI is smoothly
distributed along an axis running roughly southeast to northwest, while
HI gas of lower surface brightness extends beyond the outermost contour
shown.  The overall HI morphology at low angular resolution agrees
well with the WSRT-only images as presented in \citet{janowiecki15}.

At low and medium resolutions, the HI morphology is highlighted by
moderate HI mass surface densities throughout most of the galaxy. At
low (Figure~\ref{HIimages}a) and medium (Figure~\ref{HIimages}d)
angular resolutions, HI columns exceed N$_{\rm HI}$ $\gsim$
5\,$\times$\,10$^{20}$ cm$^{-2}$ (4.0 \msun\,pc$^{-2}$).  Two maxima in HI
column density are located on either end of the HI
distribution.  The densest HI gas (N$_{\rm HI}$ $\gsim$
9\,$\times$\,10$^{20}$ cm$^{-2}$ or 7.2 \msun\,pc$^{-2}$) is located
in the southeastern end of the HI distribution, while the secondary HI
maximum peaks at only slightly lower HI column densities.  In between
these maxima is the morphological centroid of the HI gas, where the HI
column densities are slightly lower.  

At 200 pc resolution (Figure~\ref{HIimages}g) we resolve only the
highest peaks in mass surface density in the HI gas.  Interestingly, the
HI maxima seen in the lower resolution images resolve into multiple
individual HI peaks.  At a few locations the HI column density exceeds
the ``canonical'' threshold of N$_{\rm HI}$ $=$ 10$^{21}$ cm$^{-2}$
that was empirically associated with ongoing star formation in dwarf
irregular galaxies \citep{skillman87}.  The southeastern HI maximum
seen at low and medium resolutions achieves the highest HI mass
surface density in the entire galaxy (N$_{\rm HI}$ $=$ 1.28\,$\times$\,10$^{2}$
cm$^{-21}$ or 10.3 \msun\,pc$^{-2}$).

As we discuss in detail below, the HI gas in Coma\,P is more
complicated than a simple thin disk such as those found in many other
dwarf galaxies.  With this caveat in mind, we define the morphological
center of Coma\,P by examining the medium-resolution HI data.  Using
the 3\,$\times$\,10$^{20}$ cm$^{-2}$ contour as a guide, we find that
an ellipse with an axial ratio of 2.1:1, oriented along a major axis
rotated 155\arcdeg\ east of north, provides a suitable representation
of the HI morphology at this surface density level.  The HI centroid
derived using this approach is ($\alpha$,$\delta$) = (12$^{\rm
  h}$32$^{\rm m}$10$^{\rm s}$.3,
$+$20\arcdeg25$^{\prime}$23$^{\prime\prime}$).  The corresponding
inclination is 63\arcdeg, which agrees with the $i$ $=$
6\arcdeg 3$\pm$4\arcdeg derived by \citet{janowiecki15}.

\begin{figure*}
	\begin{center}
		\includegraphics[width=0.8\textwidth]{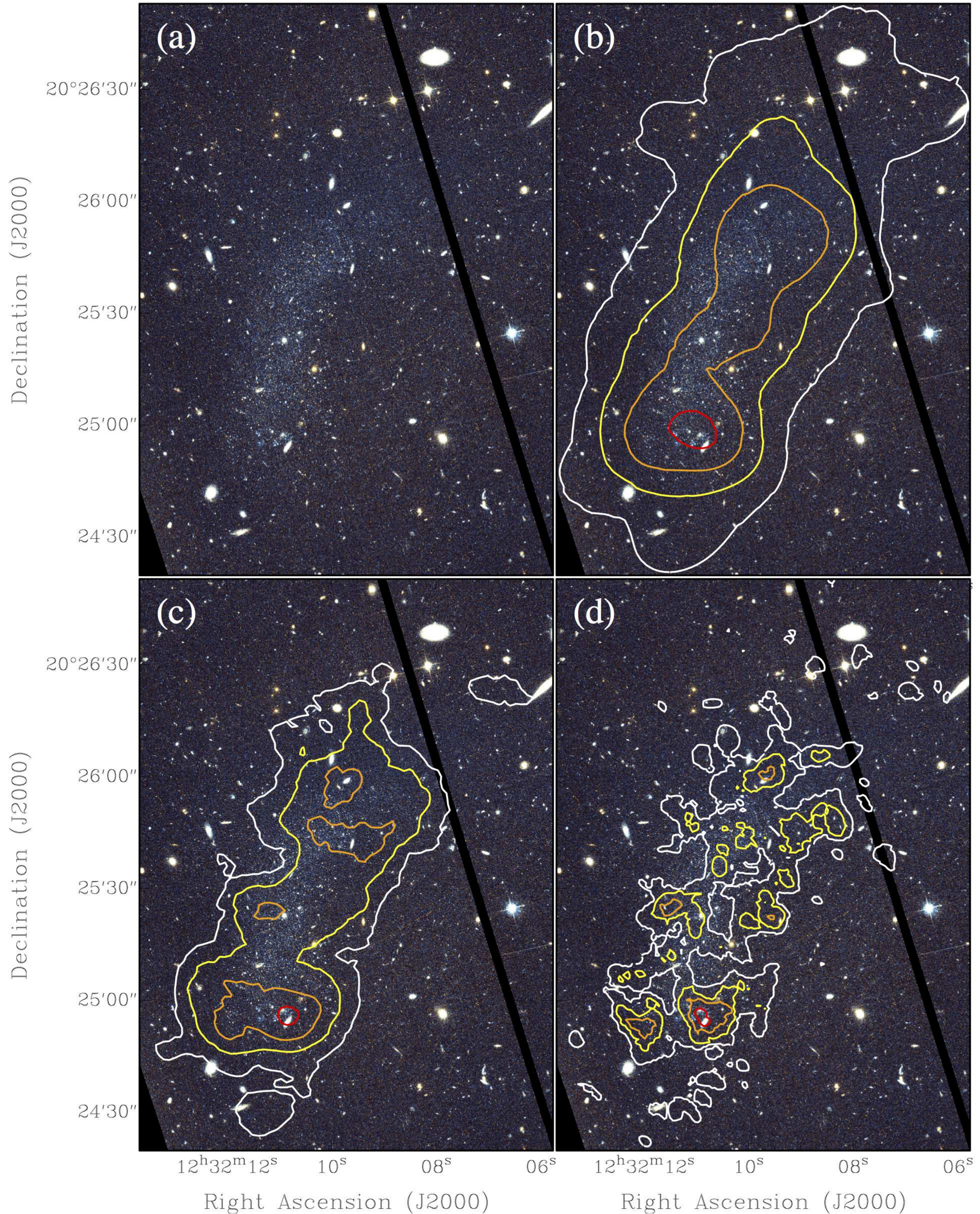}
	\end{center}
	\caption{Comparison of Hubble Space Telescope and VLA$+$WSRT HI
		imaging of Coma\,P.  Panel (a) shows the color image of the galaxy
		(see analysis in S. Brunker \etal\ 2017, in preparation) with no contours so as to
		allow detailed inspection of the stellar population and to identify
		foreground stars and background galaxies.  HI column density
		contours at low (17\arcsec\ beam), medium (12\farcsec 5 beam) and
		high (7\farcsec 5 beam) resolutions are overlaid in panels (b)--(d), respectively.  The contours are the same as shown in panels
		(a), (d), and (g) of Figure~\ref{HIimages}.  In panel (b), HI column
		densities of (2,4,6,8)\,$\times$\,10$^{20}$ cm$^{-2}$ are encoded by
		white, yellow, orange, and red contours, respectively.  In panel
		(c), HI column densities of (3,5,7,9)\,$\times$\,10$^{20}$ cm$^{-2}$
		are encoded by white, yellow, orange, and red contours,
		respectively.  In panel (d), HI column densities of
		(3,6,9,12)\,$\times$\,10$^{20}$ cm$^{-2}$ are encoded by white,
		yellow, orange, and red contours, respectively. }
	\label{HSTimages}
\end{figure*}

The HI distribution is resolved along both the major and minor axes.
We thus use the parameters derived above to create the radial profiles
of the HI gas in Coma\,P that we present in Figure~\ref{HIprofile}.
These profiles \citep[created using the \textsc{ELLINT} task in the
  \textsc{GIPSY}\footnote{The Groningen Image Processing System
    (\textsc{GIPSY}) is distributed by the Kapteyn Astronomical
    Institute, Groningen, Netherlands.} software package;][]{gipsy92}
are explicitly corrected for beam smearing using the Gaussian
approximation discussed in \citet{wang16}.  Specifically, the
corrected and uncorrected HI sizes (D$_{\rm HI}$ and D$_{\rm HI,0}$,
respectively) are related to the lengths of the major and minor axes
of the beam (B$_{\rm maj}$ and B$_{\rm min}$, respectively) via
\begin{math} {\rm D_{HI} = \sqrt{D_{HI,0}^2 - B_{maj}{\cdot}B_{min}}}\end{math}.
Beam smearing is negligible when the observed HI size is much larger
than the angular size of the beam; this is the case for our HI observations
of Coma\,P.

The radial profiles allow us to quantify both the HI size and the HI
scale length of Coma\,P.  The average (radially integrated) HI mass
surface density in the inner region is nearly flat within errors at a
value of $\sim$4.25 \msun\,pc$^{-2}$ (at 12\farcsec 5 resolution).
There is a small increase in mass surface density at $\sim$0.75 kpc in
both profiles, corresponding to where the radial integration passes
through both the primary and the secondary HI maxima.  Moving outward
the profile falls off smoothly.  The horizontal dotted lines in
Figure~\ref{HIprofile} demonstrate that the HI diameter is 4.0$\pm$0.2
kpc measured at the 1 \msun\,pc$^{-2}$ level. The HI scale length of
Coma\,P is measured by fitting an exponential function to the six data
points at radii larger than 1 kpc in the low-resolution radial profile
(Figure~\ref{HIprofile}(b)).  We find a scale length of
0.8\,$\pm$\,0.1 kpc, where the uncertainty is estimated by including
and then excluding the next inner radial profile point in the exponential
fit.

\subsection{HI Kinematics}
\label{S3.2}

The HI kinematics of Coma\,P are extremely complex. The
intensity-weighted HI velocity fields presented in panels (b), (e),
and (h) of Figure~\ref{HIimages} do not show the signatures of
coherent, solid-body rotation that are typical of low-mass galaxies
\citep[e.g.,][]{oh15}.  Rather, Coma\,P has multiple velocity
gradients within the HI distribution, along different position angles.
For example, at low resolution, in the southern region of the galaxy
there is a smooth velocity gradient spanning $\sim$20 \kms\ oriented
along the HI major axis.  In the central region of the HI
distribution, gas at the highest recessional velocity spans the entire
minor-axis diameter of the galaxy.  Moving northward, the isovelocity
contours become nearly perpendicular to those in the southern region.
A simple scenario of a smoothly rotating disk is insufficient to
describe the global kinematics.  This conclusion was already evident
in the WSRT data as presented by \citet{janowiecki15}.  We note that
our low-resolution velocity field is consistent with the one shown in
that work.

\begin{figure*}
	\begin{center}
		\includegraphics[width=0.75\textwidth]{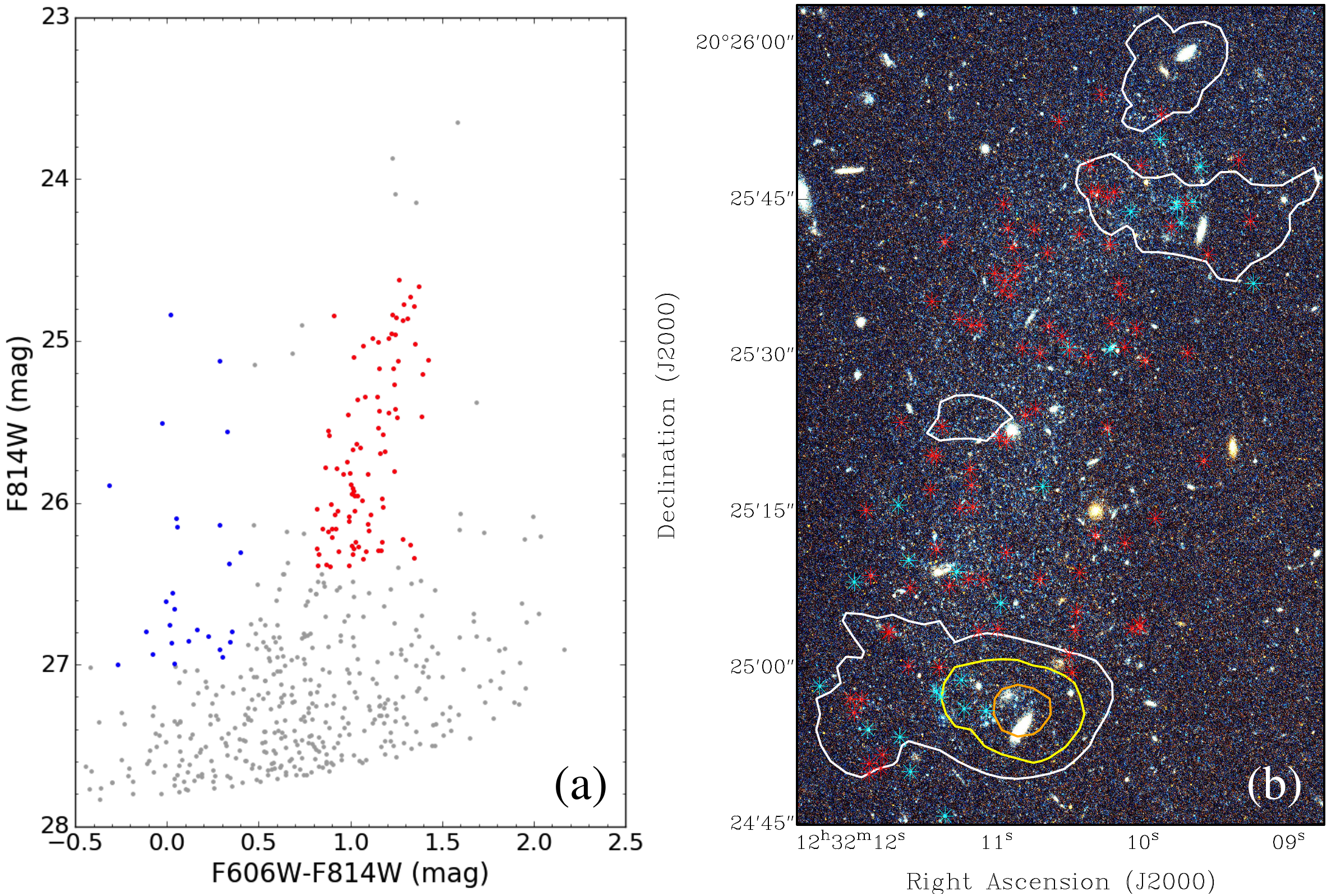}
	\end{center}
	\caption{The locations of red and blue stars in Coma\,P.  Panel (a)
		shows the color-magnitude diagram as presented in Brunker \etal\ (2017, in
		preparation).  Selected red and blue stars are color-coded (see
		detailed discussion in \S~\ref{S4}).  These stars are plotted as
		cyan (representing blue stars) and red (representing red stars)
		asterisks in panel (b). The color HST image is slightly enlarged compared to the field shown in Figure~\ref{HSTimages}.  HI column
		density contours from the medium-resolution (12\farcsec 5 beam) data
		are overlaid: white, yellow, and orange encode HI column densities
		of (7, 8, 9)\,$\times$\,10$^{20}$ cm$^{-2}$, respectively.  The red
		stars are uniformly distributed throughout the galaxy, while the
		blue stars are strongly concentrated in the regions of highest HI
		mass surface density. }
	\label{RedBluestars}
\end{figure*}

The images with higher angular resolution show similar kinematic signatures
to those seen in the low-resolution velocity field.  At 333 pc
resolution (Figure~\ref{HIimages}(e)), the southern velocity
gradient and the changing orientations of the isovelocity contours in
the north are still present. The lower surface brightness sensitivity
of these images results in some noticeable changes, especially in the
north and central regions.  At the highest angular resolution we again 
are sensitive to only the densest parcels of gas.  Thus for clarity we
do not overlay isovelocity contours on Figure~\ref{HIimages}(h). The signatures of extremely complicated HI gas kinematics remain prominent at all angular resolutions.

The intensity-weighted velocity dispersion images of Coma\,P are shown
in panels (c), (f), and (i) of Figure~\ref{HIimages}.  At low angular
resolution the magnitudes of the HI velocity dispersion range from
$\sim$6 to $\sim$12 \kms, typical of the values found
in low-mass galaxies \citep[e.g.,][]{mcnichols16}.  In the southern
region of the galaxy (where the nominal smooth velocity gradient was
identified in the discussion above) the velocity dispersions remain
below 10 \kms.  In the center of the HI distribution the values
increase to $\sim$12 \kms, corresponding to the onset of the
changes in directions of the isovelocity contours.

Moving to the medium- and high-resolution images shown in panels (f)
and (i) of Figure~\ref{HIimages}, the same signatures remain: lower
velocity dispersions are found in the south, with a significant
increase toward the north.  Interestingly, the northern and southern
maxima in HI mass surface density have very different dispersion
characteristics.  The gas with higher mass surface density in the south is
characterized by low dispersion gas (v $\simeq$ 8 \kms), with a small
increase to $\sim$10 \kms\ on the far eastern portion of the HI
maximum.  In contrast, the northern HI peak (which contains
sub-structure at 12\farcsec 5 resolution) contains the region of the
galaxy with gas of the highest dispersion (v $\simeq$ 14 \kms).

It is important to note that the collapse of a three-dimensional datacube
into a two-dimensional image loses information as a result of the weighting by
intensity per pixel.  A comparison of Figure~\ref{HIimages} with the
channel maps shown in Figures~\ref{CHMAPS.LOW}, \ref{CHMAPS.MED}, and
\ref{CHMAPS.HIGH} demonstrates this by the difference in velocity
ranges between the two- and the three-dimensional products.  Given the
kinematic complexity of Coma\,P, we therefore explore the full
three-dimensional data products in \S~\ref{S5} below.
\section{HI and Star Formation in Coma\,P}
\label{S4}

To understand the relationship of the HI gas to the underlying stellar
populations in Coma\,P, in Figure~\ref{HSTimages} we overlay the same
sets of HI column density contours as shown in panels (a), (d), and
(g) of Figure~\ref{HIimages} on a color HST image (created using the
F606W filter image as blue, the F814W filter as red, and a linear
average of the two filters as green; see S. Brunker \etal\ 2017, in
preparation). Figure~\ref{HSTimages}(a) shows the same HST
color image as in S. Brunker \etal\ (2017, in preparation), with no HI contours so as
to allow detailed inspection of the stellar population and to identify
foreground stars and background galaxies.  As described in detail in
\citet{janowiecki15} and S. Brunker \etal\ (2017, in preparation), the stellar
population of Coma\,P has a remarkably low surface brightness (peak
surface brightnesses in the g$^{\prime}$, r$^{\prime}$, and
i$^{\prime}$ bands of 26.4$\pm$0.1 mag\,arcsec$^{-2}$, 26.5$\pm$0.1
mag\,arcsec$^{-2}$, and 26.1$\pm$0.1 mag\,arcsec$^{-2}$,
respectively).  With a morphology similar to that of the HI images
shown in Figure~\ref{HIimages}, the stellar population arcs from
southeast to northwest.  Nearly all of the compact and luminous
sources seen in this field are foreground stars or background
galaxies.  The individual stars in Coma\,P are faint but spatially
resolved, resulting in the diffuse distribution of stars that is
contained within the HI contours.  A few stellar clusters are
apparent, and these are discussed in further detail below.

In panels (b)-(d) of Figure~\ref{HSTimages}, the HI contours
as shown in Figure~\ref{HIimages} are color-coded for ease of
interpretation.  Comparing the low- and medium-resolution HI contours
to the HST images, we find excellent spatial agreement between the HI gas of
high column density and the locations of the faint but resolved
stars in Coma\,P.  Effectively all of the stellar population of Coma\,P
is located within HI gas with N$_{\rm HI}$ $\gsim$
5\,$\times$\,10$^{20}$ cm$^{-2}$ (4.0 \msun\,pc$^{-2}$).  The highest
HI mass surface densities in the southeastern region of the galaxy are
cospatial with what appear to be individual stellar clusters or
extended star formation regions.

To further investigate the relationship between the HI gas and the
resolved stars in Coma\,P, Figure~\ref{RedBluestars} shows the
locations of individual red and blue stars on the HST color image.
Panel (a) shows the color-magnitude diagram as presented in S. Brunker
\etal\ (2017, in preparation).  Therein, blue and red stars are color-coded.
Blue stars are selected as having F814W magnitudes brighter than 27.0
and F606W$-$F814W colors bluer than 0.4, while red stars are the
individual red giants used for the distance determination as discussed
in S. Brunker \etal\ (2017, in preparation).  These red and blue stars are then
plotted on the color HST image in panel (b) as cyan and red asterisks.
HI column density contours from the medium-resolution
(12\farcsec 5 beam) image are shown at levels of (7, 8,
9)\,$\times$\,10$^{20}$ cm$^{-2}$.  As expected for an old stellar
population that is well mixed, the red stars are uniformly distributed
across the galaxy.  In contrast, the blue stars are decidedly
concentrated within the regions of highest HI mass surface density.

\begin{figure*}
	\begin{center}
		\includegraphics[width=0.75\textwidth]{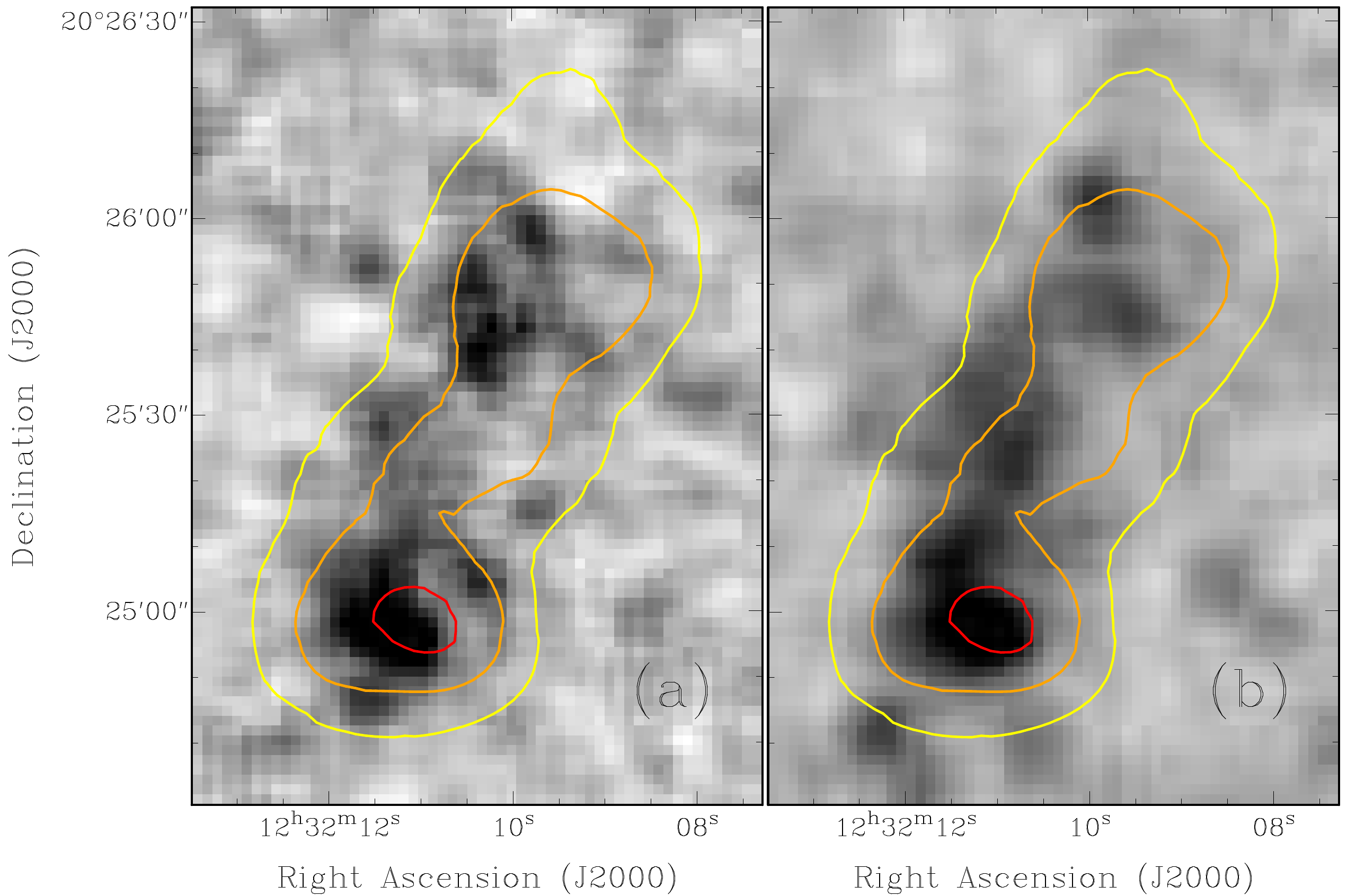}
	\end{center}
	\caption{Comparison of GALEX and VLA$+$WSRT HI imaging of Coma\,P.
		Panel (a) shows the far ultraviolet image and panel (b) shows the
		near ultraviolet image. Each image has been smoothed by a Gaussian
		kernel to the GALEX angular resolution (4\farcsec 2 and
		5\farcsec 3 in the near and far ultraviolet, respectively).  Low-resolution (17\arcsec\ beam) HI column density contours (see also
		Figures~\ref{HIimages} and \ref{HSTimages}) at levels of
		(4,6,8)\,$\times$\,10$^{20}$ cm$^{-2}$ are encoded by yellow,
		orange, and red contours, respectively.  The ultraviolet emission
		detected by GALEX is cospatial with the HI gas detected in Coma\,P.
		Note that the southern maximum in HI column density is cospatial with
		the maxima in both GALEX images, although there is a background
		galaxy near this position (see Figure~\ref{HSTimages}).}
	\label{GALEXimages}
\end{figure*}

As discussed in detail in \citet{janowiecki15}, the star formation
properties of Coma\,P are extreme.  The source is not detected in deep
ground-based \halpha\ imaging, implying that the current rate of massive star
formation is zero.  Curiously, Coma\,P is detected by GALEX in
both the near-UV (NUV) and far-UV (FUV) bands (although at low S/N).  We compare the low-resolution HI image with the GALEX FUV and NUV images in
Figure~\ref{GALEXimages}. The ultraviolet emission detected by GALEX
is cospatial with the HI gas detected in Coma\,P.  By comparison with
the HST images in Figure~\ref{HSTimages} it is evident that the flux
from the resolved stellar populations is dominated by stars that are
capable of producing UV photons but not of photoionizing HII regions
(S. Brunker \etal\ 2017, in preparation).  Note that the southern maximum in HI column
density is cospatial with the maxima in both GALEX
images. This is the same location identified above in the HST images
as harboring stellar clusters.  We note, however, that there is also a
significant background spiral galaxy nearby that makes this
interpretation difficult (see the highest HI contour in
Figure~\ref{HSTimages}(c) and in both panels of Figure~\ref{GALEXimages}).

We estimate the FUV-based star formation rate using the calibration
derived from resolved stars in \citet{mcquinn15b}.  The resulting
global far-ultraviolet star formation rate is very low: 
SFR$_{\rm FUV}$ $=$ (3.1$\pm$1.8)\,$\times$\,10$^{-4}$ \msun\,yr$^{-1}$.
This is equal to the lowest far-ultraviolet star formation rate
amongst the galaxies in the SHIELD sample \citep{teich16}.  However,
the FUV flux in Coma\,P is extended over a much larger area than in
the (physically smaller) SHIELD galaxies.  A spatially resolved
analysis of the FUV star formation rate density versus the HI mass
surface density \citep[e.g., as presented in][]{teich16} is not
possible at the low S/N ratio of the GALEX FUV images.  Deeper UV
imaging of Coma\,P would be especially useful for studying the star
formation on timescales of 100-200 Myr.
\section{The Complicated HI Dynamics of Coma\,P}
\label{S5}
\begin{figure*}
	\begin{center}
		\includegraphics[width=0.8\textwidth]{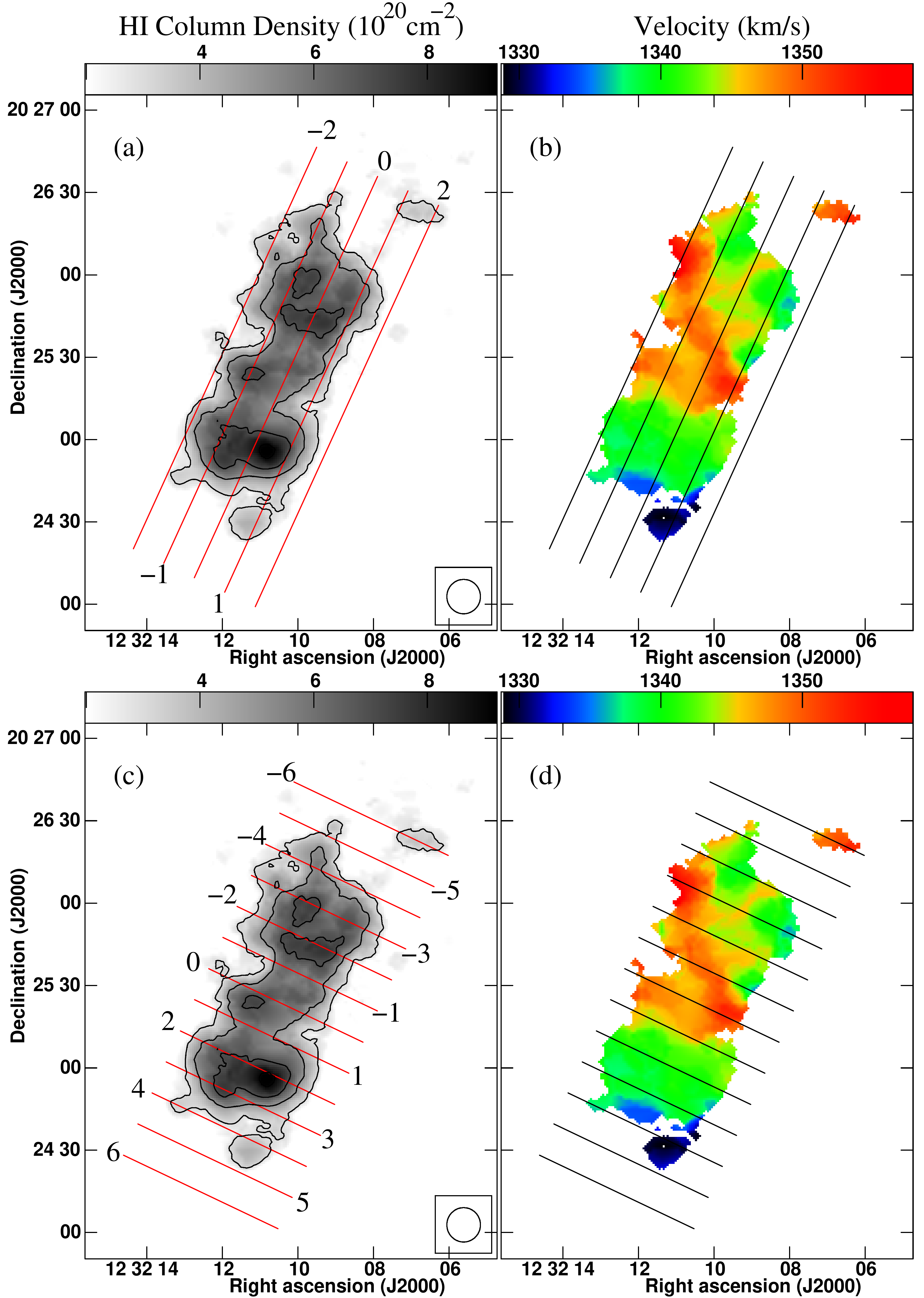}
	\end{center}
	\caption{Spatially resolved position-velocity analysis of Coma\,P.
		Medium-resolution (12\farcsec 5 beam) HI column density images
		(panels (a) and (c)) and intensity-weighted velocity fields (panels (b) and
		(d)) are shown. Overlaid are the positions of the major axis (panels
		(a) and (b)) and minor axis (panels (c) and (d)) position-velocity slices.
		Each slice is the width of the HI beam, is separated by the beam
		width, and is numbered for reference with respect to the resulting
		slices as shown in Figures~\ref{majorslices} and \ref{minorslices}.
		The central major and minor axis slices cross at the adopted
		dynamical center of the galaxy ($\alpha$,$\delta$) = (12$^{\rm
			h}$32$^{\rm m}$10.$^{\rm s}$3,
		$+$20\arcdeg25$^{\prime}$23$^{\prime \prime}$).  }
	\label{slices}
\end{figure*}
As discussed in \S~\ref{S3.2}, the HI kinematics of Coma\,P are not
amenable to standard modeling techniques. The two- and three-dimensional data products (see Figures~\ref{CHMAPS.LOW},
\ref{CHMAPS.MED}, \ref{CHMAPS.HIGH}, \ref{HIimages}) demonstrate this
clearly. In an effort to understand the complex neutral gas dynamics
of Coma\,P, we thus perform two types of analysis on the three-dimensional
datacubes.

\subsection{Spatially Resolved Position-Velocity Analysis}
\label{S5.1}

In the first line of investigation we apply the spatially resolved
position velocity analysis developed in \citet{cannon11b} and employed
in \citet{mcnichols16} and \citet{cannon16}.  This technique allows us
to estimate the projected rotation velocity of a source when its
kinematics are not well described by a simple tilted ring.  The
approach relies on identifying a projected HI major axis, which is
typically that axis along which the largest velocity gradient is
present.  In some cases this HI major axis aligns with the major axis
of the stellar component.  However, this is not necessarily the case
for all dwarf galaxies, especially for those systems with extremely
unusual gas kinematics \citep[see application of the approach to the
  post-starburst galaxy DDO\,165 in][]{cannon11b}. The HI minor axis
is assumed to be offset from the major axis by 90\arcdeg.
\begin{figure*}
	\begin{center}
	\includegraphics[width=0.67\textwidth]{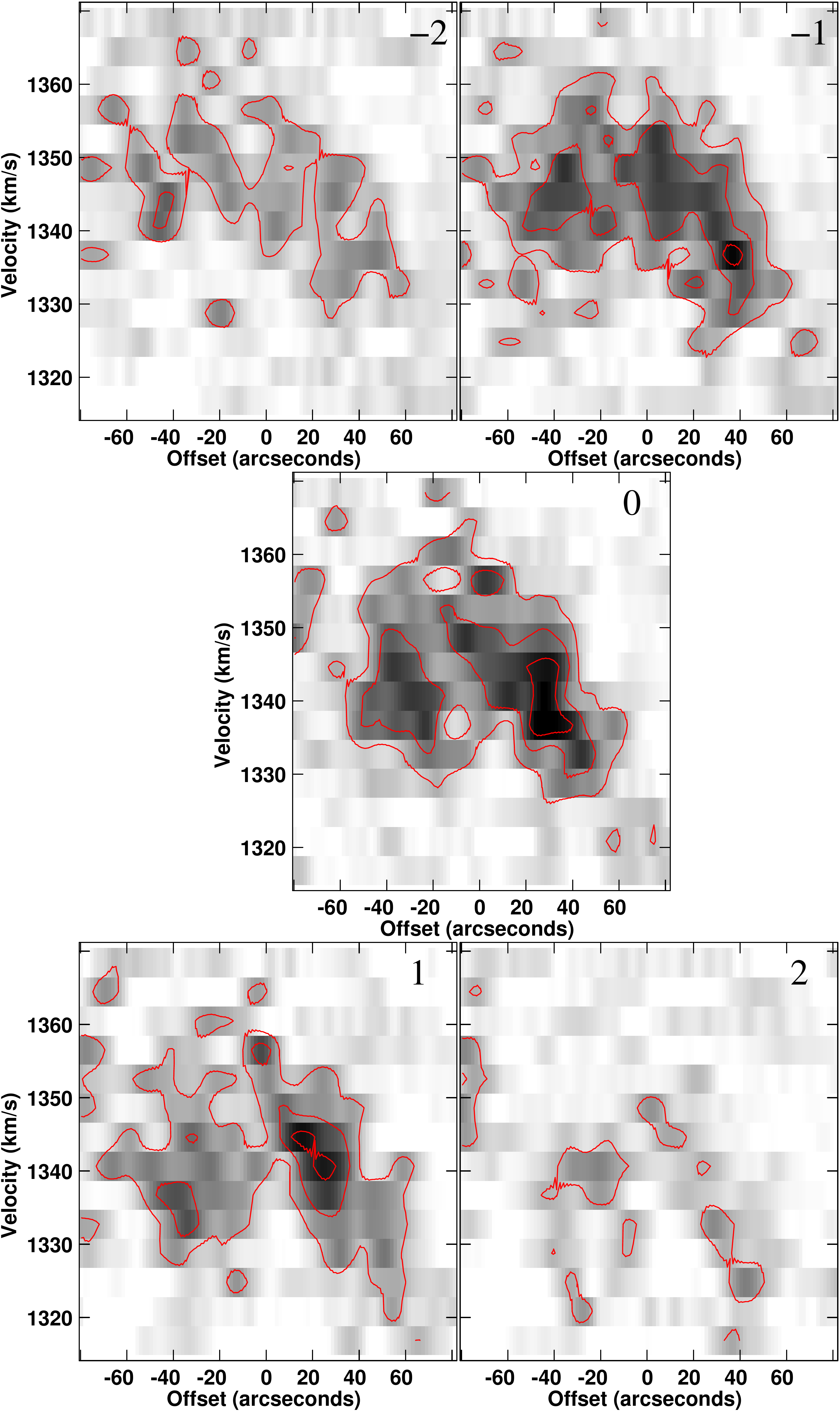}
\end{center}
\caption{Position-velocity slices for the major axis through the
	medium-resolution (12\farcsec 5 beam) data cube for Coma\,P.  The
	slices are numbered as shown in Figure~\ref{slices}. Slices with
	negative numbers are east of the central slice, while slices with
	positive numbers are west of it.  Positive angular
	offsets correspond to moving from the center of the slice toward the
	southeast, while negative angular offsets correspond to moving from
	the center of the slice toward the northwest.  Contours are overlaid
	at the 3$\sigma$, 6$\sigma$, 9$\sigma$, and 12$\sigma$ levels.}
\label{majorslices}
\end{figure*}
The spatially resolved position analysis then examines a
collection of position-velocity slices for the major and minor axes.  Each
slice is the width of, and is offset from the neighboring slice by,
the HI beam width.  The series of major and minor axis slices allow
the identification of the largest projected rotation velocity, both
along the major axis and also as a change in velocity of the HI
centroids of the minor axis slices.  Further, this approach allows us
to isolate kinematically distinct features \citep[e.g.,][]{cannon11b}
in the gas.

To balance angular resolution with sensitivity, in this analysis we
examine the medium-resolution (12\farcsec 5 beam) data cube (see
Figure~\ref{CHMAPS.MED}).  We identify the HI major axis interactively
by examining position velocity slices at a range of position angles.
The resulting major axis is oriented at a position angle of
335\arcdeg\ measured east of north.  This angle is in agreement with
the direction of the maximum projected velocity gradient in the moment
one images at both low and medium resolutions (see
Figure~\ref{HIimages}), and it is also in rough agreement with the
position angle of the stellar component (see Figure~\ref{ComaPfield}).
The HI minor axis is 90\arcdeg\ away (at 65\arcdeg), again rotated east
through north.  The HI dynamical center is assumed to be cospatial
with the adopted morphological center (see discussion in
\S~\ref{S3.1}): 
($\alpha$,$\delta$) = (12$^{\rm h}$32$^{\rm m}$10$^{\rm s}$.3, 
$+$20\arcdeg25$^{\prime}$23$^{\prime\prime}$).
Note that since the slices are the width of the HI beam, the absolute
position of the dynamical center is not critical for interpretation.

\begin{figure*}
	\begin{center}
		\includegraphics[width=0.67\textwidth]{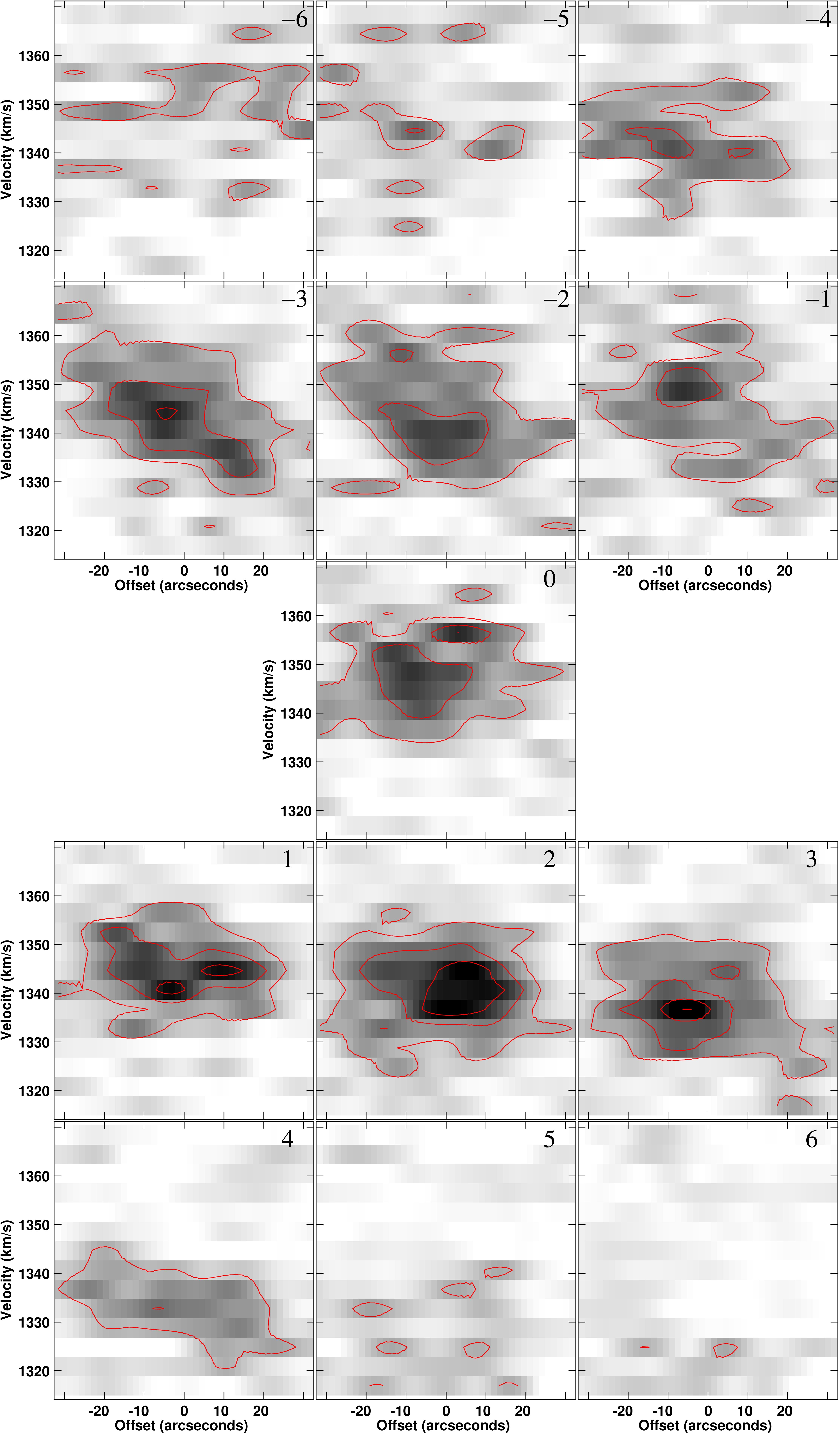}
	\end{center}\vspace{-0.5 cm}
	\caption{Position-velocity slices for the minor axis through the
		medium-resolution (12\farcsec 5 beam) data cube for Coma\,P.  The
		slices are numbered as shown in Figure~\ref{slices}. Slices with
		negative numbers are north of the central slice, while slices with
		positive numbers are south of it. Positive angular
		offsets correspond to moving from the center of the slice toward the
		southwest, while negative angular offsets correspond to moving from
		the center of the slice toward the northeast.  Contours are overlaid
		at the 3\,$\sigma$, 6\,$\sigma$, 9\,$\sigma$, and 12\,$\sigma$ levels.}
	\label{minorslices}
\end{figure*}

Based on the 12\farcsec 5 (333 pc) beam, we then symmetrically slice
(i.e., use the same number of slices at positive and negative angular
offsets from the major and minor axes) the data cube along both the
major and the minor axes as shown in Figure~\ref{slices}.  Five major
axis slices are required to cover the HI distribution of Coma\,P,
while 13 minor axis slices are required.  Such an axial ratio is
expected based on the inclination derived above.  On both the major
and the minor axes, the central slice is numbered ``0.''  For the
major axis slices, those with negative numbers are located east of the
central slice and those with positive numbers are located west of the
central slice.  Likewise, for the minor axis slices, those with
negative numbers are located north of the central slice and those with
positive numbers are located south of it.  

The resulting spatially resolved position-velocity slices for the major and minor axes are presented in Figures~\ref{majorslices} and
\ref{minorslices}, respectively.  In each figure the axes and color
scalings are identical.  Contours are overlaid at the 3$\sigma$,
6$\sigma$, 9$\sigma$, and 12$\sigma$ levels in each panel.  These
figures contain a wealth of information about the kinematics of the HI
gas in Coma\,P.

Beginning with the major axis slices shown in
Figure~\ref{majorslices}, all slices show two opposing velocity
gradients: a crescent-shaped structure in the plane of velocity versus offset, the shape and dynamic range of which are the same as shown in
\citet{janowiecki15}.  This signature of two opposing velocity
gradients is most easily interpreted as two separate velocity
components within the HI distribution of Coma\,P.  Moving from the
central slice to either positive or negative offset reveals that the
HI velocity structure on the major axis does not change as a function
of offset from the center.  This confirms the identification of the
major axis of one of the HI components.  Further, it demonstrates that
the turnover in velocity as a function of offset, likely caused by the
presence of a second HI kinematic component, extends throughout the
entire HI distribution of Coma\,P.

The minor axis slices confirm the interpretation of two kinematic
components in the HI gas.  The HI velocity structure along the minor
axis does indeed show a moving velocity centroid from one side of the
galaxy to the center (e.g., examine the highest S/N contours in the
series of slices with negative numbers).  However, importantly, around
the central slice this signature turns over in velocity space and then
continues in the opposite direction on the other side of the galaxy
(examine the highest S/N contours in the series of slices with
positive numbers).  Such a turnover in velocity space is very unusual.
A galaxy with normal HI kinematics (that are well described by a
simple tilted ring model) would show the signature of changing centroid
position across all of the minor axis cuts.  Note that even if the
inclination were in fact very low (unlikely for Coma\,P based on the
optical and HI morphologies) and the signature of projected rotation
very subtle, it would still not result in a turnover in velocity space 
as a function of position.

Using the spatially resolved position velocity analysis as a guide, we
can coarsely separate the two HI gas components in the observed low-
and medium-resolution datacubes.  We stress that this separation is
not clean, either spatially or spectrally. HI gas connects these two
kinematic components in the data cubes and in the position velocity
slices through them.  With this caveat in mind, we estimate that the
relative HI mass ratio of the two HI components is in the range of
(0.6:1) to (0.8:1); varying the masking and threshold levels used to
identify the two components changes the resulting mass ratio.  While
the absolute value of the mass ratio is uncertain, the masses of the two components
are of the same order.

It is difficult to interpret the velocity gradients seen in the
position velocity slices of Figures~\ref{majorslices} and
\ref{minorslices} as coherent rotation.  If we take the projected
velocity gradients at face value, then we can make coarse estimates of
the dynamical masses of the two kinematic components.  We use the
turnover in projected velocity along the central major axis slice as
the boundary between the two components.  The more massive component
(extending to positive angular offsets in Figure~\ref{majorslices})
has a maximum observed projected velocity gradient of $\Delta v$ $=$
35\,$\pm$\,10 \kms\ that spans $\sim$60\arcsec.  Likewise, the less
massive component also spans an angular offset of
$\sim$60\arcsec\ with a maximum observed projected velocity gradient
of $v =$ 30\,$\pm$\,10 \kms.  If we interpret each of these maximum
projected velocity gradients as the signature of rotation of each
kinematic component, then the inclination-corrected rotational
velocities of the more and less massive components are
$\sim$20\,$\pm$\,10 \kms\ and $\sim$17\,$\pm$\,10 \kms, respectively;
each subtends $\sim$800 pc (30\arcsec). We stress that this
interpretation is highly uncertain, since the dynamics of Coma\,P are
certainly a mixture of rotation and of non-circular motions.

Using the estimates of angular offset and rotational velocity above, a
simple estimate of the dynamical masses (M$_{\rm dyn}$
$=$ \begin{math} \frac{V^2{\cdot}R}{G}\end{math}) of the two kinematic
components of Coma\,P results in M$_{\rm dyn}$ $=$
7.2\,$\times$\,10$^{7}$ \msun\ and 5.3\,$\times$\,10$^{7}$ \msun.  We
conservatively estimate a 50\% uncertainty to highlight the
aforementioned difficulties in determining the physical properties of
Coma\,P.  Correcting the HI gas mass by a factor of 1.35 for helium,
metals and molecules, and assuming M$_{\star}$ $=$
(1.0$\pm$0.3)\,$\times$\,10$^{6}$ \msun\ (see S. Brunker \etal\ 2017, in
preparation), we find a total baryonic mass M$_{\rm bary}$ $=$
4.8\,$\times$\,10$^{7}$ \msun.  Summing the dynamical masses of the
two components (M$_{\rm dyn}$ $\simeq$ 1.2\,$\times$\,10$^{8}$ \msun)
within the extent of the HI shows that Coma\,P is dominated by dark matter at the
level of roughly 2.5 to one.


\subsection{Three-Dimensional Modeling of the HI Distribution}
\label{S5.2}
\begin{figure*}
	\begin{center}
		\includegraphics[width=0.8\textwidth]{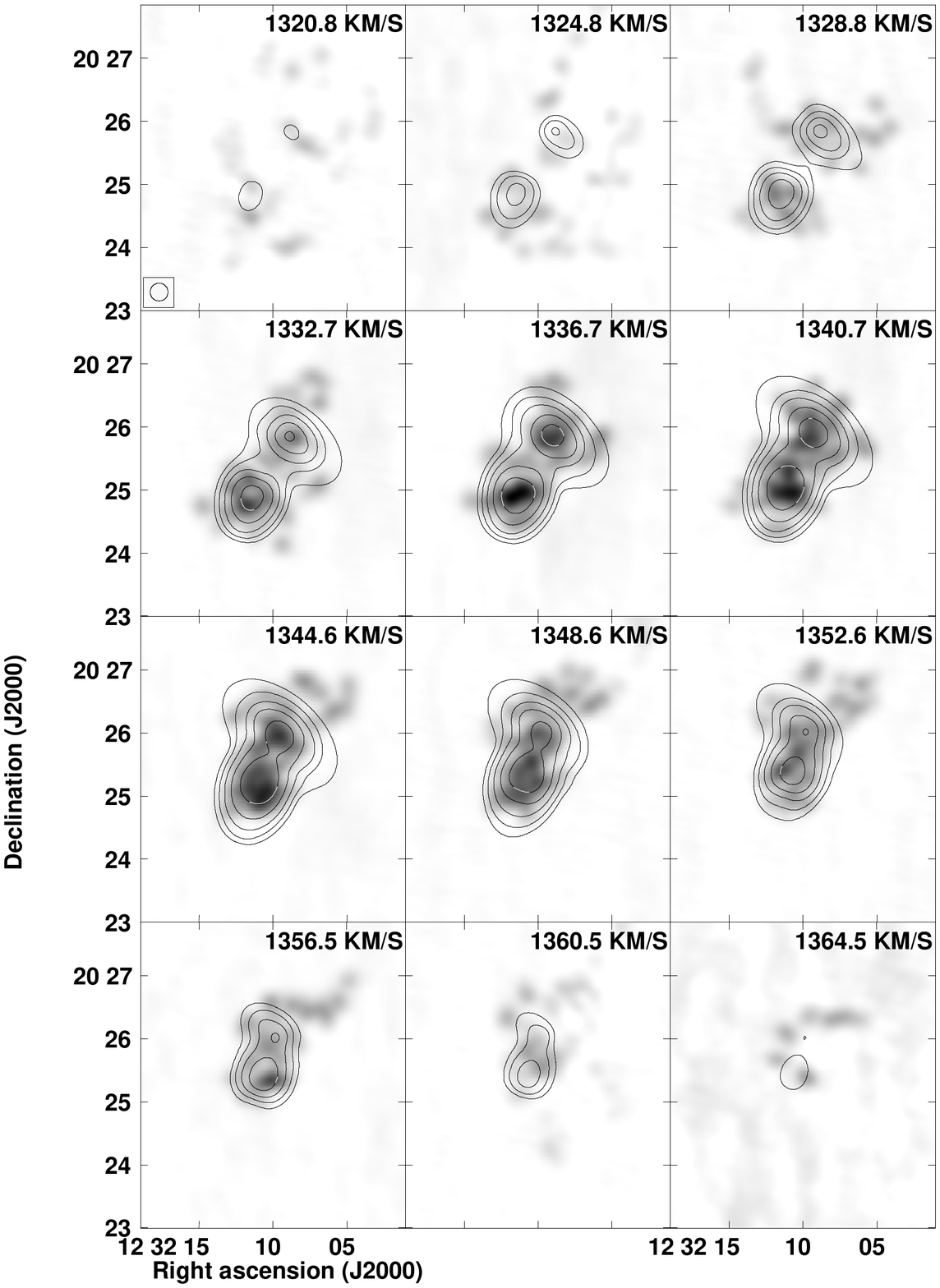}
	\end{center}
	\caption{Comparison of observation (grayscale) and TiRiFiC 3D model
		(contours) for Coma\,P.  The grayscale channel maps are the same as
		shown in Figure~\ref{CHMAPS.LOW}. The beam size is 17\arcsec.  The
		best-fit TiRiFiC model consists of two kinematic components with a
		relative HI mass ratio of 0.64:1.  The model contours are spaced by
		a factor of two and highlight the full dynamic range of the model
		cube.  There is good agreement between model and observation,
		demonstrating that the complex kinematics of Coma\,P are adequately
		described by two separate HI components that are experiencing either
		a collision or an infall episode.}
	\label{tirific.low}
\end{figure*}

\begin{figure*}
	\begin{center}
		\includegraphics[width=0.8\textwidth]{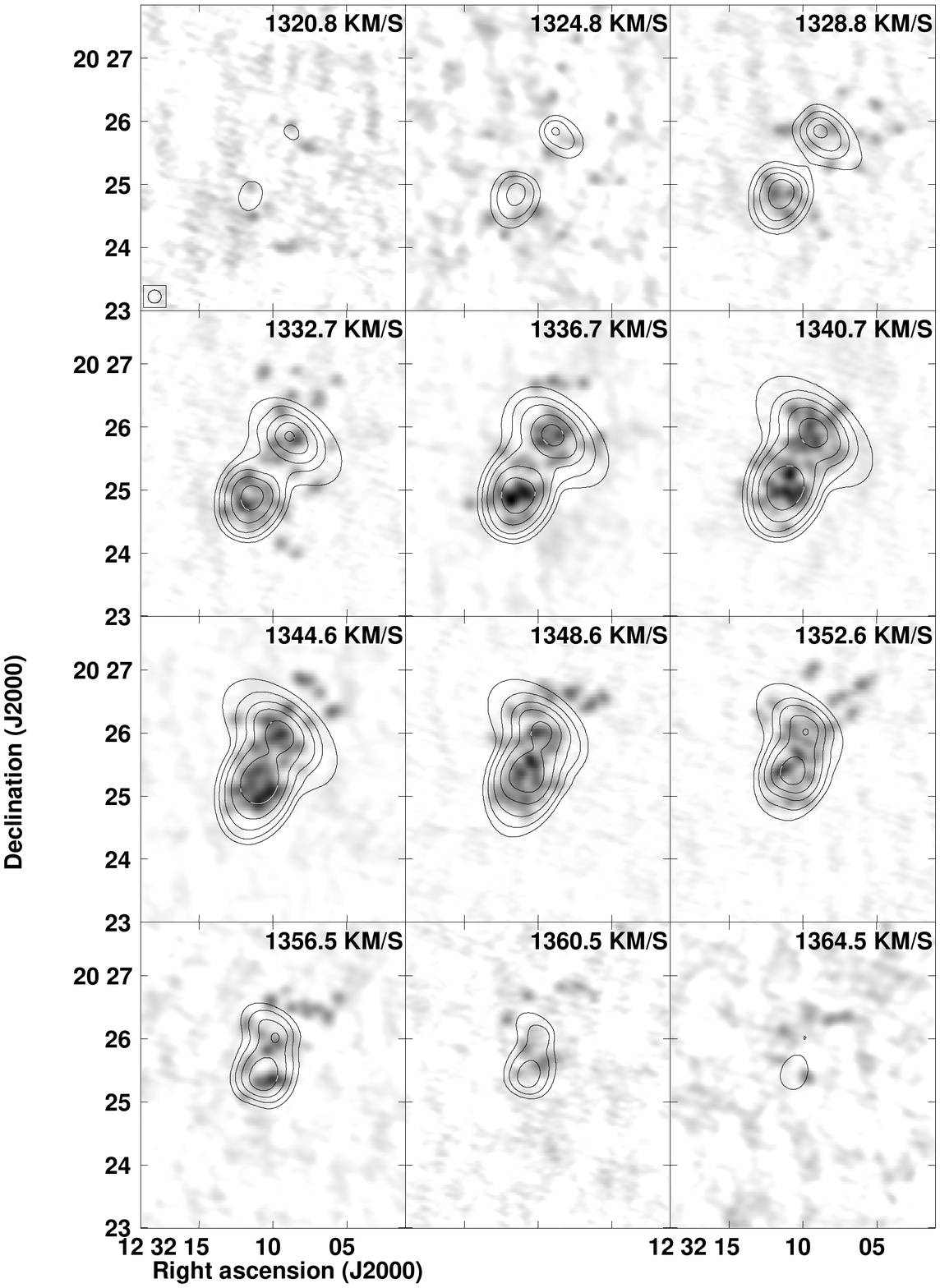}
	\end{center}
	\caption{Comparison of observation (grayscale) and TiRiFiC 3D model
		(contours) for Coma\,P.  The grayscale channel maps are the same as
		shown in Figure~\ref{CHMAPS.MED}. The beam size is 12\farcsec 5.  The
		best-fit TiRiFiC model consists of two kinematic components with a
		relative HI mass ratio of 0.64:1.  The model contours are spaced by
		a factor of two and highlight the full dynamic range of the model
		cube.  There is good agreement between model and observation,
		demonstrating that the complex kinematics of Coma\,P are adequately
		described by two separate HI components that are experiencing either
		a collision or an infall episode.}
	\label{tirific.med}
\end{figure*}

The second line of investigation attempts to fit a three-dimensional
model to the HI data cube using the ``Tilted Ring Fitting Code''
(``TiRiFiC'') software package \citep{jozsa07}.  By working with the
full three-dimensional data cube, TiRiFiC eliminates the loss of
information when using only the velocity field for tilted ring
fitting.  Further, TiRiFiC allows for modeling of multiple HI disks
and for inhomogeneities in the surface brightness distribution.  In
situations where complicated gas kinematics are present (as in
Coma\,P), this allows a direct comparison of the resulting model with
the observed data cube.

The TiRiFiC package was used in interactive mode to fit the medium-resolution (12\farcsec 5 beam) data cube.  The resulting best-fit
models contain two HI components: a more massive component (located in
the southeastern region of the galaxy) and a less massive segment that
is undergoing either infall or outflow and is moving with respect to
the center of the galaxy.  No simple model with a single inclined disk
was able to accurately describe the HI kinematics of Coma\,P.

Figures~\ref{tirific.low} and \ref{tirific.med} present this best-fit
model as contours overlaid on the same low-resolution and
medium-resolution channel maps as shown in Figures~\ref{CHMAPS.LOW}
and Figures~\ref{CHMAPS.MED}, respectively.  The model reproduces the
complicated three-dimensional gas distribution and kinematics very
well.  Moving from low to high velocities, one can easily differentiate
the two HI components, both spatially and spectrally.  The opposing
velocity gradients of the two components are also clearly visible.
The two model kinematic components have a relative mass ratio of 0.64:
1, with the more massive component in the southeastern region of
the disk.  This compares favorably with the HI and dynamical mass
ratios of the two components estimated in \S~\ref{S5.1}.

The interpretation of the best-fit model from TiRiFiC is very similar
to the interpretation of the spatially resolved position velocity
slices discussed above.  Specifically, the best-fit model contains two
kinematically distinct HI components.  The kinematic discontinuity in
the central region of the galaxy (highest velocity gas in
Figures~\ref{tirific.low} and \ref{tirific.med}) is the result of the
presence of two HI components of differing masses.  Whether this event
is best described by two low-mass galaxy disks or by the infall of HI
gas into Coma\,P is not certain.  Given the current and recent
quiescence of the galaxy in terms of star formation rate (see
discussion above and in S. Brunker \etal\ 2017, in preparation), it is unlikely
that the kinematics are described as an outflow driven by star-formation.
The extended gas of low surface brightness would then be the result of
this ongoing interaction or infall event.

\section{Contextualizing Coma\,P}
\label{S6}

\begin{figure*}
	\begin{center}
	\includegraphics[width=0.75\textwidth]{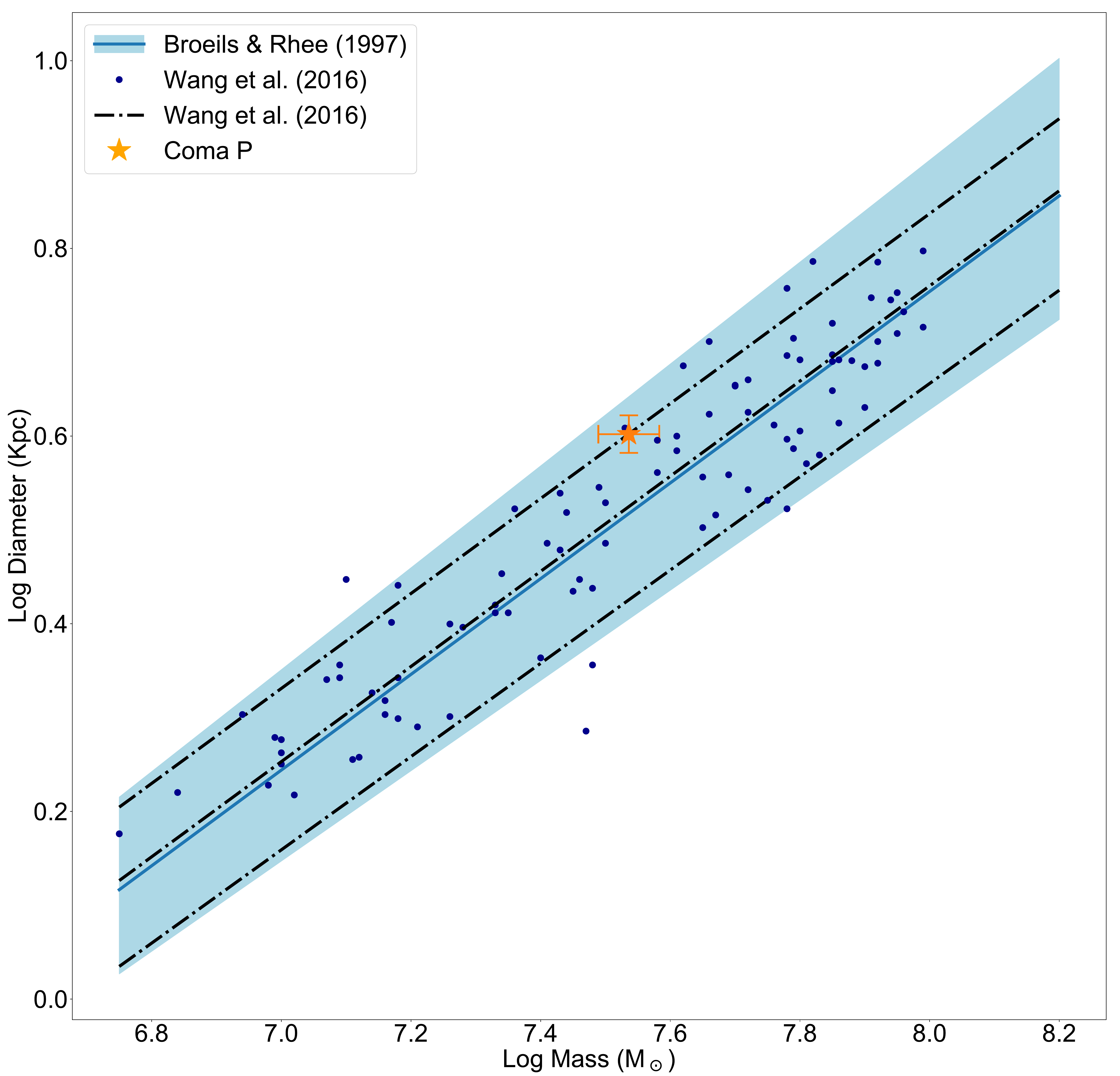}
	\end{center}
	\caption{The M$_{\rm HI}$ -- D$_{\rm HI}$ relation, showing Coma\,P
		and a subset of the comparison galaxies from \citet{wang16}.  The HI
		diameters of all galaxies are determined at the 1 \msun\,pc$^{-2}$
		level.  The dash--dot lines show the best-fit relation and the
		3\,$\sigma$ scatter from \citet{broeils97}. The solid line shows the
		best-fit relation from \citet{wang16}, and the shaded blue region
		shows the 3\,$\sigma$ scatter.  The slopes of the M$_{\rm HI}$ --
		D$_{\rm HI}$ relation as derived by \citet{broeils97} and
		\citet{wang16} are effectively identical.  Coma\,P is just
		consistent with the 3\,$\sigma$ scatter of the relation of \citet{wang16}. This galaxy can be considered to have a large HI diameter
		given its HI mass or a low HI mass given its physical size. }
	\label{MHI-DHI}
\end{figure*}
Various observed properties of Coma\,P make it an extremely unusual
dwarf galaxy when compared to other objects in the Local Volume.  The
proximity of the galaxy is unusual given its large recessional
velocity.  The source is currently quiescent in terms of massive star
formation but yet has significant UV flux indicative of star formation
on timescales of 100-200 Myr.  The optical surface brightness is
extremely low, and the HI properties are unusual for a low-mass halo.
We now attempt to contextualize Coma\,P with respect to other Local
Volume galaxies.

Comparing Coma\,P to the low-mass galaxies in the SHIELD sample
\citep{cannon11a,mcnichols16,teich16}, Coma\,P is slightly more
HI-massive than all but one of these galaxies.  Given this HI mass,
Coma\,P is large in physical size.  As discussed in \S~\ref{S3.1}, the
HI diameter is measured to be 4.0\,$\pm$\,0.2 kpc at the 1
\msun\,pc$^{-2}$ level.  This allows us to place Coma\,P on the
M$_{\rm HI}$--D$_{\rm HI}$ relation--the empirical relationship
between HI size and total HI mass \citep[see first characterization
  in][]{broeils97}.  While the origins of this relation are likely
multi-faceted, \citet{wang16} use data for hundreds of nearby galaxies
to show that the relationship is linear over four orders of magnitude
in HI mass (7 $\lsim$ log(M$_{\rm HI}$/M$_{\odot}$) $\lsim$ 11). The
relation appears to continue to lower masses, although the small
number of sources makes this extension uncertain at the present time.

In Figure~\ref{MHI-DHI} we show the M$_{\rm HI}$--D$_{\rm HI}$
relation including Coma\,P and a subset of the data from the sample of
\citet{wang16}.  Coma\,P is just consistent with the 3$\sigma$
fit derived by \citet{wang16} (note that the parameterizations of the
M$_{\rm HI}$--D$_{\rm HI}$ relation by \citet{broeils97} and \citet{wang16} are
effectively identical).  The galaxy can be considered to have a large
HI distribution given its HI mass, a smaller total HI mass given its
physical size, or an artificially extended HI size due to the presence
of two HI components.

\begin{figure*}
	\begin{center}
	\includegraphics[width=0.75\textwidth]{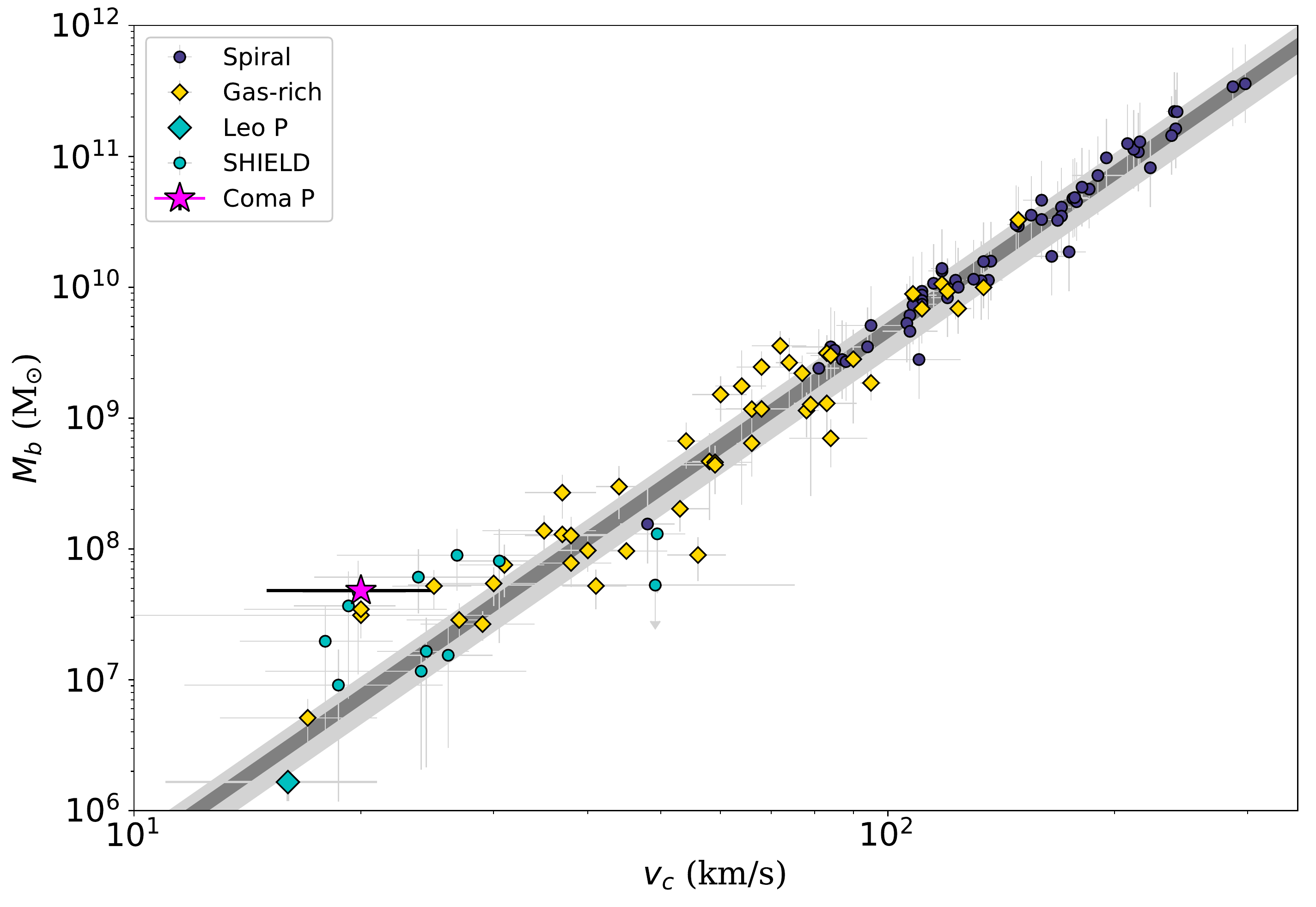}
	\end{center}
	\caption{The Baryonic Tully--Fisher relation (BTFR) as derived in
		\citet{mcnichols16} and \citet{cannon16}.  Coma\,P is shown by a
		magenta symbol.  We note that the position of Coma\,P in this plot
		is demonstrative only.  The rotational velocity is that of the more
		massive dynamical component only, while the baryonic mass is the
		total for the entire Coma\,P system.  The dark and light shaded gray
		regions represent the 1\,$\sigma$ and the 3\,$\sigma$ deviations
		from a fit of the BTFR to the gas-rich galaxy sample, respectively.
		Coma\,P lies just off of this calibration of the BTFR.}
	\label{BTFR}
\end{figure*}

It is important to stress that the M$_{\rm HI}$--D$_{\rm HI}$ relation
is meaningful for isolated galaxy disks where the physical size of the
HI component is directly related to the gravitational potential of the
parent halo.  A complicated system such as Coma\,P, where the HI
kinematics favor a model with multiple HI components, may not be
appropriately interpreted in the context of the 
M$_{\rm HI}$--D$_{\rm HI}$ relation.  With this caveat in mind, taken
at face value, the HI is considerably extended in Coma\,P compared to
expectations based on the total HI mass of the system.


In addition to the large HI size of Coma\,P, its total HI mass is
large compared to its stellar mass. \citet{janowiecki15} estimate a
very large M$_{\rm HI}$/M$_{\star}$ ratio using stellar masses derived
from standard mass-to-light scaling relations. While 
M$_{\rm HI}$/M$_{\star}$ is independent of distance, S. Brunker \etal\ (2017, in
preparation) suggest a significantly revised stellar mass from HST
imaging of M$_{\star}$ $=$ (1.0$\pm$0.3)$\times$10$^6$ \msun.  Regardless
of the exact value of M$_{\star}$, Coma\,P is one of the most HI-rich
galaxies known relative to its stellar mass, especially among sources
with accurately measured M$_{\rm HI}$/M$_{\star}$.  The estimated
change in stellar mass moves Coma\,P within the scatter of the sources
with stellar masses near 10$^6$ \msun\ in \citet{huang12} and
\citet{janowiecki15}.  However, we note that this is not a completely
fair comparison, since these M$_{\rm HI}$/M$_{\star}$ ratios are less
well measured (deep optical observations tend to recover more flux,
thus reducing the estimated M$_{\rm HI}$/M$_{\star}$ ratio). The
revised M$_{\rm HI}$/M$_{\star}$ value also places Coma\,P slightly
above the extrapolation of the best-fit trend line for low surface
brightness galaxies found in \citet{mcgaugh18}.


Additionally, we note that, especially at low mass, other individual
systems have been identified with higher M$_{\rm HI}$/M$_{\star}$
ratios than the one that we find in Coma\,P.  For example,
\citet{matsuoka12} find that the northeastern component of the system
HI\,1225$+$01 is more extreme than Coma\,P, although Coma\,P is
of lower surface brightness.  Similarly, some of the recently
discovered resolved stellar populations of the UCHVC population discovered by
ALFALFA \citep{adams13,adams16} have M$_{\rm HI}$/M$_{\star}$ ratios
in excess of 100 \citep{janesh15,janesh17}.

Coma\,P is shown on the baryonic Tully--Fisher relation
\citep{tullyfisher77} in Figure~\ref{BTFR}.  The derivations of
rotational velocity and baryonic mass use the same techniques for
Coma\,P (\S~\ref{S5.1}) and the low-mass comparison samples
\citep{ezbc14,mcnichols16}.  However, importantly, we stress that the
dynamics of Coma\,P favor a multiple-disk scenario, and therefore it
can only be shown on this plot with the estimate of rotation velocity 
from one of the two kinematic components.  We use the inclination-corrected 
velocity of the more massive component ($v
\sim$20\,$\pm$\,10 \kms) and the total baryonic mass of the entire
Coma\,P system (which is well represented by the gas mass because the
stellar mass is so low).  The position of Coma\,P is highly uncertain
and is meant to be demonstrative only.

Compared to the SHIELD galaxies and other systems studied in
\citet{mcnichols16}, Coma\,P is slightly offset from the best-fit
relation in the plane of M$_{\rm bary}$ versus V$_{\rm c}$, although in
agreement within the errors on the measurement of the rotational
velocity.  However, we stress that the error bars on the estimates of
the rotational velocity of Coma\,P make its placement representative
only.  Interestingly, while Coma\,P lies just slightly above the
empirical threshold differentiating rotationally-supported and
pressure-supported systems that was identified by \citet{mcnichols16},
its interpretation is nonetheless very challenging due to its
complicated HI gas dynamics.

\section{Conclusions}
\label{S7}

We have presented new HI spectral line imaging of the curious Local
Volume dwarf irregular galaxy Coma\,P (AGC\,229385). This source is
unique in many respects.  It has an extremely low optical surface
brightness, complicated HI gas dynamics, and very intriguing patterns
of recent star formation.  An understanding of the physical properties
of this galaxy is important for galaxy evolution in the low-mass
regime.

This galaxy is one of three HI line sources detected by ALFALFA at
recessional velocities between 1282 and 1343 km\,s$^{-1}$ and in close
angular proximity to each other.  Remarkably, the new HST imaging presented in
S. Brunker \etal\ (2017, in preparation) has shown that Coma\,P is much closer
than the original Virgocentric flow model of \citet{masters05}
predicted (D $\simeq$ 25 Mpc).  At D $=$ 5.50\,$\pm$\,0.28 Mpc, the
total HI mass is M$_{\rm HI}$ $=$
(3.48$\pm$\,0.35)\,$\times$\,10$^{7}$ \msun, reshaping and
complicating our interpretive framework for this already enigmatic
source.  At this time the physical association of Coma\,P with the
other HI line sources (AGC\,229384 and AGC\,229383) remains
unconfirmed, although based on their angular and velocity proximity it
is likely that they are all located at roughly the same distance.

Using a combination of deep WSRT and VLA observations, we study the HI
properties of Coma\,P at physical resolutions of $\sim$450, 333, and
200 pc.  The HI morphology is characterized by two maxima in mass surface
density near either end of an inclined HI distribution.  The HI
disk of Coma\,P is physically large (diameter 4.0$\pm$0.2 kpc measured
at the 1 \msun\,pc$^{-2}$ level) compared to galaxies of comparable HI
mass.  Its globally averaged HI surface density is slightly lower than
expected based on the known scaling between HI mass and HI
diameter.  At the same time, it has an exceptionally high
  ratio of HI mass [M$_{\rm HI}$ $=$
  (3.48$\pm$0.35)\,$\times$\,10$^{7}$ \msun] to stellar mass
  [M$_{\star}$ $=$ (1.0$\pm$0.3)\,$\times$\,10$^{6}$ \msun].

An intrinsically faint dwarf, Coma\,P's extreme optical surface
brightness renders it completely undetectable in SDSS.  Coma\,P is
currently not forming massive stars (as traced by nebular
\halpha\ emission), even though the HI mass surface densities exceed
10$^{21}$ cm$^{-2}$ in localized regions.  The galaxy has formed stars
during the most recent few hundred million years.  Far-ultraviolet emission as
traced by GALEX images shows good spatial agreement with the high
column density HI gas in the inner regions of the galaxy.

The HI kinematics of Coma\,P are extremely complex and are not
well described by a simple tilted ring model.  There are highly
irregular gas motions and macroscopic velocity gradients at multiple
position angles.  Using both three-dimensional modeling and spatially
resolved position velocity analysis, we find that the HI properties are
best described either by two colliding HI disks or by a significant
infall event.  A coarse separation of these HI components shows that
they are of comparable mass.

Using the most obvious signatures of rotation in the three-dimensional
data, we estimate the total dynamical mass of the Coma\,P system to be
(1.2\,$\pm$\,0.6)\,$\times$\,10$^{8}$ \msun.  It is important to note
that this is the sum of the estimates of the dynamical masses of the
two kinematically distinct HI components.  Coma\,P is not unusually
dominated by dark matter (M$_{\rm dyn}$/M$_{\rm bary}$ $\simeq$ 2.5)
compared to other Local Volume dwarf galaxies with similar dynamical
masses.  In the context of the fundamental M$_{\rm HI}$--D$_{\rm HI}$
and baryonic Tully--Fisher relations, Coma\,P is slightly offset from,
but in general agreement with, the trends seen for other Local Volume
galaxies.  We stress that the presence of two HI components
complicates the interpretation of these fundamental scaling relations.

The origins of the exceptional physical properties of Coma\,P may
result from one or more infall and/or interaction events.  There are
HI clouds with similar velocities to that of Coma\,P in close angular
proximity.  While we do not detect stellar components of these clouds
in deep ground-based images \citep{janowiecki15}, if these are in fact
physically associated with Coma\,P then they may suggest a past
significant merger event.  We refer the reader to the discussions in S. Brunker
\etal\ (2017, in preparation) of the local environment and large-scale
 structure surrounding Coma\,P.

Assuming that the irregular kinematics of the HI gas in Coma\,P do
result from a merger of two low-mass disks, then this offers a unique
opportunity to compare observations with predictions from simulations.
Hydrodynamical simulations with realistic feedback prescriptions are
just now able to model halos in the mass range of Coma\,P (see, for
example, the discussion of the ``Feedback in Realistic Environments''
simulations that are discussed in S. Brunker \etal\ 2017, in preparation).  The
unusual gas kinematics in Coma\,P are qualitatively similar to those
shown in the simulations of gas-rich dwarf galaxies and dark
companions by \citet{starkenburg16}.  Further, those simulations
naturally predict off-center star formation in the resulting merged
system - similar to the properties of HI gas and star formation seen
in Coma\,P.

Regardless of their origins, the physical characteristics of Coma\,P
make it an exceptional galaxy.  Its physical properties are among the
most extreme of any known source in the Local Volume.  Coma\,P will
serve as a critical benchmark for our understanding of low-mass galaxy
evolution.

\acknowledgements

J.M.C. and C.B. acknowledge support from past NSF grant AST-1211683.
The authors are grateful to Macalester College for supporting this
project.  The ALFALFA team at Cornell is supported by NSF grants AST-
0607007 and AST-1107390 to R.G. and M.P.H. and by grants from the
Brinson Foundation.  S.J. acknowledges support from the Australian
Research Council's Discovery Project funding scheme (DP150101734).
K.L.R. and W.F.J. are supported by NSF grant AST-1615483.
M.G.J. acknowledges support from the grant AYA2015-65973-C3-1-R
(MINECO/FEDER, UE).

\facility{VLA, HST, GALEX, WIYN 3.5m}



\clearpage
   

\clearpage


\clearpage


\end{document}